%% file: survey.tex
\documentclass[10pt,journal]{IEEEtran}
\let\IEEEtransubparagraph\subparagraph
\let\subparagraph\paragraph

\usepackage{color}

\ifCLASSINFOpdf
\else
\fi
\usepackage[hyphens]{url}

\usepackage{titlesec,graphicx,enumitem,indentfirst,caption,tabularx,multirow,booktabs,lipsum,multicol}

\let\subparagraph\IEEEtransubparagraph
\titlespacing*{\section}
{0pt}{1.0ex }{1.2ex}
\titlespacing*{\subsection}
{0pt}{1.0ex }{1.2ex}
\titlespacing*{\subsubsection}
{0pt}{1.0ex }{1.2ex}

\begin{document}
%
\title{Privacy Leakage in Mobile Computing: Tools, Methods, and Characteristics}
%
%
%

%

\author{{\fontsize{12pt}{15pt}\selectfont Muhammad~Haris,{ The~Hong~Kong University~of~Science~and~Technology}\\  Hamed~Haddadi, {Queen~Mary~University~of~London and Qatar Computing Research Institute}\\
Pan~Hui,{ The~Hong~Kong University~of~Science~and~Technology}}\\
{\fontsize{10pt}{13pt}\selectfont \textit{ (mhmughees, panhui)@cse.ust.hk, and hamed@eecs.qmul.ac.uk }}}

\setlength\extrarowheight{5pt}

\maketitle

\begin{abstract}
\input{abstract}
\end{abstract}


%
\IEEEpeerreviewmaketitle

\input{intro}

\input{mobilecomp}

\input{conclusion}

\twocolumn[
  \begin{@twocolumnfalse}
    \maketitle

    \begin{appendix}
  \label{appn}
    \end{appendix}
  \end{@twocolumnfalse}
]

\begin{table}[httbp]
\begin{center}

\begin{tabularx}{\textwidth}{| p{2.2cm}|p{2.85cm}|p{2.1cm}|p{2.5cm}|p{6.30cm}|}
\hline
\hline

\textbf{Authors/Study}    & \textbf{Year - Number of Users } & \textbf{Feature
Surveyed}& \textbf{Platform  } & \textbf{Summary } \\ \hline

Christopher Thompson et al.~\cite{thompson2013s} &	2013 - 189 &	Sources of 
Information leak	& Android +iOS &
17\% users don’t understand that background applications have same ability as foreground. 
Many users misunderstood about source of information leaks.
\\ \hline
felt et al.~\cite{felt2012android} & 2012 - 308 &	Effectiveness of Android permissions &	Android & Only 
17\% users pay attention to permissions during app installation.
Moreover, only 3\% users understand scope and implication of permissions. Permissions of android are not complete failure nor complete success. Users' privacy related opinions should also be shared with application permissions at installation.
\\ \hline

Kelley et al~\cite{kelley2012conundrum} &
2012 - 20 &	Effectiveness of Android permissions &	Android &
Participant do not understand the terms used in permissions notifications.
They highly depend on ratings, word of mouth, and reviews.
Additionally, they are not aware of of threats and malware applications in android market.
\\ \hline
Balebako et al~\cite{balebako2013little} &
2013 - 19 &	Perception  on Privacy Leak &	Android &
13 participants were unaware that data can be shared for advertisement purpose.
They were also unaware of scope of data sharing in terms of frequency and location of data.
\\ \hline
Chin et al.~\cite{chin2012measuring} &
2012 - 60 &	Perception on security and privacy &	iOS+ Android &
Users are less willing to perform sensitive task (like banking) on their mobile phones than there laptops. They are also more concerned about privacy on their mobile devices than their laptops.
Reason for these concerns are: Physical device lost, User interface concerns and
Misconceptions of network connection.

\\ \hline
Kyoung et al.~\cite{choe2013nudging} &
2013 - 129 &	Effectiveness of visual privacy alerts or framing effect & Android &	Results suggest that visual representations of privacy information of apps can influence installation decisions by smartphone users. Majority of participants commented that they found the privacy rating very helpful in deciding whether to install an app ,when it was disclosed visually on the installation page of the app. 

\\
\hline

King et al.~\cite{king2013come} &	2013 - 13 &	User Expectation about mobile privacy & Android+iOS &	Current privacy controls do not full fill users expectations. Many users believe that variety of assurance structures(such as developers reputation) protect them from privacy leaks. Users rank SMS, email and photos more sensitive in terms of privacy than locations information.
\\ \hline
Benenson et al.~\cite{benenson2013should} &	2013 - 506 &	Privacy risk Perception &	Android + iOS &
iOS users are very less aware of privacy and sensitive data types on their devices.
Similarly they do not have much concerned about security, while android users are usually more active in installing security software on their devices.
\\ \hline
Barkhuus et al.~\cite{barkhuus2003location} &	2003 - 16 &	Privacy concerns on location based services &	Multiple &
Privacy concerns for location tracking are much higher than position aware services. However, if users find a service useful than they are willing to share their exact location to tracking service also. 
\\ \hline
Braunstein et al.~\cite{braunstein2011indirect}	& 
2011 - 200 &	Indirect privacy survey &	Multiple &
Asking users directly about privacy is not accurate measure of their privacy attitude as it makes them think about potential privacy risks explicitly. However indirect privacy survey’s are more better measurement.

\\ \hline

\end{tabularx}
\end{center}

\end{table}

\begin{table*}[h]
\begin{center}

\vspace{-30em}
\begin{tabularx}{\textwidth}{| p{2.2cm}|p{2.85cm}|p{2.1cm}|p{2.5cm}|p{6.30cm}|}

\hline 
Kelley et al.~\cite{kelley2013privacy}	&
2013 - 20 &	Understand How user selects apps &	Android &
Users do not understand privacy permissions displayed to them during app installation.
It is suggested that privacy implications should be included in the main page of applications and must be clear and simple.

\\ \hline
Shklovski et al.~\cite{shklovski2014leakiness} &	2014 - 187 &	User concerns about privacy &	Android+ ios &
Users concerns about privacy are over ridden by other factors while installing apps.
They have misconceptions about what data apps can access on their devices. They are also concerned about unnecessary data leaked to third part businesses. It is found that 58\% individuals in the study have previously deleted apps due to privacy concerns.

\\ \hline
Felt et al.~\cite{felt2012ve} &	2012 - 115 &	Rank users concerns about mobile devices resources & Multiple &
Warnings in iOS and android do not consider users concerns. Locations sharing is mid-level risk, users are more concerned about contacts, sms, emails, photos, calls and calendars etc. Moreover they rank particular data type less private if they have controls to monitor it themselves e.g turn of location. In addition, they rank risks involving third parties higher.

\\ \hline  
Häkkilä et al.~\cite{hakkila2005s} &
2005 - 119 &	User perception about privacy & 	Multiple &
85\% users consider their mobile phones very private device. They regard text messages more private than emails. Since text message are not secured survey reveals users expectation on using encryption or any other security means to protect their SMS.

\\ \hline
Benisch et al.~\cite{benisch2011capturing}  &
2010 - 27 &	User preferences of privacy	& Symbian &
93\% users are comfortable sharing their locations with family and friends, 60\% of them with facebook friends, 57\% with university community and 36\% with advertisers. Its is also found that users have privacy preferences with multiple dimensions such as when to share, what to share and with whom to share.

\\ \hline
Muslukhov et al.~\cite{muslukhov2012understanding} &
2011 - 22 &	User understanding to security &	Multiple &
Users store sensitive data on their phone and concerned about it, however they do not take any action for security of their data. Pin codes based and password lock security measures are unusable for them. They believe that current privacy protection mechanisms require to much effort from them.

\\
\hline 
\end{tabularx}
\end{center}
\centering  

\end{table*}

\clearpage
\bibliography{haris}
\bibliographystyle{IEEEtran}

\input{biographies}

\end{document}

%% file: abstract.tex
The number of smartphones, tablets, sensors, and connected wearable devices are rapidly increasing. Today, in many parts of the globe, the penetration of mobile computers has overtaken the number of traditional personal computers. This trend and the \textit{always-on} nature of these devices have resulted in increasing concerns over the intrusive nature of these devices and the privacy risks that they impose on users or those associated with them. In this paper, we survey the current state of the art on mobile computing research, focusing on privacy risks and data leakage effects. We then discuss a number of methods, recommendations, and ongoing research in limiting the privacy leakages and associated risks by mobile computing.  

%% file: intro.tex
\section{\textbf{Introduction}}
\label{sec:intro}

\IEEEPARstart{T}{oday}, we are surrounded by a powerful combination of mobile technology and Internet connectivity. Many of these devices have become our personal assistants and connectivity gateway to the world. In the past few years, reduced manufacturing costs and advances in hardware technologies (e.g. sensors, processors) and software platforms (e.g. Android, iOS)  have made smartphones increasingly more powerful and popular. 

More recently, a new family of mobile devices called wearable technology have also appeared in the technology scene. These devices are designed to be worn on the human body in an always-on and ubiquitous manner. Some of these devices have been designed to perform dedicated tasks with the help of cloud service or mobile phones. However, more sophisticated devices like Google Glass,\footnote{\url{https://www.google.com/glass/start/}} not only have connectivity and computational capability much like mobile phones, but also due to their contextual use and spacial sensing capabilities, they have a much broader effect on the individuals who are engaging in using them and those around them.
 
In this survey we explore and survey the potential privacy leakages in mobile computing and wearable devices. Our main contribution is to classify the leakage of privacy and also provide short summary of the multiple efforts to study, model and reduce privacy issues in mobile and wearable devices. We provide insights into current solutions for preserving privacy in these devices, on various levels.


The rest of this survey is organised as follows: in Section~\ref{S:MobileComp} we provide a brief introduction to Mobile Computing. In Section~\ref{S:CharPriv} we provide a summary of conventional methods used today for characterizing privacy. Section~\ref{S:leakage} presents a summary of current research on privacy leakage, firstly through mobile devices, then through apps in Section~\ref{S:apps}, and lastly mobile advertizing platforms in Section~\ref{S:ads}. In Section~\ref{S:sensing} we discuss mobile sensing methodologies and their privacy implications. In Section~\ref{S:users} we investigate the characteristics of individuals' behaviour and their attitude towards privacy in mobile computing. Finally, we conclude the paper in Section~\ref{S:conc} and suggest future work in the area based on the works in current survey.

%% file: mobilecomp.tex
\section{\textbf{Mobile Computing}}
\label{S:MobileComp}

In this section we provide a brief introduction to mobile computing. Our intention is not to survey the current advances in mobile computing itself, but to provide adequate fundamental information that will help the reader to understand the details of privacy leakages within the scope of this survey. Specific aspects of mobile computing and personal mobile devices include: being personal (Not shared), persistent network connectivity, and mobility (location independence). Advances in technology have redefined the meaning of mobile computing. Today, mobile computing encapsulates sophisticated smartphones that are equipped with large processing powers and various kinds of intelligent sensors. These devices have operating systems, Internet connectivity, and can run advanced applications. 

Recently new kinds of mobile devices have also emerged. These devices are similar to their handheld counterparts (i.e., smartphones) in features, but users can usually wear these on their bodies. Additionally, they are designed to seamlessly interact with the environment and are continuously \textit{connected}. The most advanced form of these devices are smart glasses. They run advanced mobile operating systems and possess computational capability. Throughout this survey, we use mobile computing to represent smartphones as well as advanced wearable computers and devices.

\section{\textbf{Characterizing Privacy in Mobile Computing}}
\label{S:CharPriv}

\subsection {Privacy: Definition}

Over the past decade, privacy has gained significant attention in academia as well as in industry. The main reason behind this interest is the consequences of privacy violation on individuals. On the one hand, sensitive user data can be exploited by malicious identities to steal or expose personal information about the users and on the other hand it can be misused to harm users financially or socially. Moreover, companies can also use this data to learn sensitive personal identifiable information about users without their consent and awareness~\cite{christin2011survey}.

Although details of private information can vary, meanings of privacy have similarities across different contexts. Many definitions are proposed by the research community to understand the social meaning of privacy. Probably the most precise explanation of privacy is by Clarke et al.~\cite{Clarke:1999:IPC:293411.293475}. They explains that {\em Privacy is the interest that individuals have in sustaining a `personal space', free from interference by other people and organizations}. Another effective meaning of privacy is provided by Westin~\cite{Westin:2003}. He views privacy as {\em the claim of individuals to determine for themselves when, how, and to what extent information about them is communicated to others}.

Privacy is a social notion with many facets. However it is also related to a wide variety of computing technologies. As suggested by Langheinrich et al.~\cite{Langheinrich02privacyinvasions}, privacy has different goals in different contexts, owing to which there can be no standard definition of privacy in mobile technology. The following definitions are provided by researchers:
\begin{itemize}[noitemsep,topsep=0pt,parsep=0pt,partopsep=0pt]
  \item
	 Montjoye et al.~\cite{de2013unique} researchers explain privacy as the guarantee that participants maintain control over the release of sensitive information that relate to them. This includes the protection of information that can be inferred from both the sensor readings themselves and from the interaction of the users with the participatory sensing system.
  \item
	Shmatikov et al.~\cite{Shmatikov02defininganonymity} provide the view of privacy in multi agent systems. They express that privacy is preserved if no malicious agent can use the system to learn how other agents make identity-based decisions.
   \item Lucas et al. in~\cite{Introna00privacyand}, define privacy as:
\begin{itemize}[noitemsep,topsep=0pt,parsep=0pt,partopsep=0pt]
  \item
	No access to an individual's personal data without informed consent;
 \item
	Individual's control over personal information;
 \item
	Freedom of the individual from judgment by others.
\end{itemize}
\end{itemize}

In order to cater to mobile and wearable devices, we define privacy through the following characteristics.
\subsection {Individual Consent}
 Considering the nature of mobile and wearable devices, individuals are device owners as well as the people surrounding the device. Consent here means the degree of agreement between the user's awareness about data collection and the actual data handling by the application. Many surveys have been carried out to estimate user expectation and awareness with respect to privacy leaks through their devices~\cite{sadeh2009understanding,braunstein2011indirect}.

\subsection {Private data}
Private data is a piece of information on mobile devices that includes user privacy. It may consist of any sensitive data, which can be exploited to get the  identity or other personal information about the user. Examples of such data may include location coordinates, device ID, contacts, pictures, video, audio etc~\cite{krishnamurthy2008characterizing}. Additionally, it is also possible that data is not directly sensitive, but analysis can be done on it to infer personal information. For example, recently many attempts have been made to use data mining techniques to infer personal information from raw data streams of various mobile sensors~\cite{chittaranjan2013mining,mohan2008nericell,lu2010jigsaw,choudhury2008mobile}.

\subsection {Prevention: Control and Transparency}
Prevention in mobile computing is the balance between privacy and functionality. Privacy leak prevention in mobile computing mainly includes anonymity, control and transparency. This has recently been debated in the \textbf{Privacy-by-Design} framework~\cite{pbd2014}.

 \textbf{Anonymity} is the measure of extent, to which certain data collected and stored by mobile applications can be linked to the identity of the individuals. Many techniques have been proposed by the research community to guarantee data anonymity~\cite{zhou2008brief,ghinita2007fast,ghinita2008anonymization,laurila2012mobile}.

\textbf{Control} is the authority of users over data collected about them. In mobile computing it means that decisions about the collection, storage and analysis of data are made only by the owners of data. This may also include control over the removal of data previously collected. Google has recently provided users authority over their own search data and allowed them to remove all data collected about them~\cite{bugiel2013flexible,fenske2012biometrics}.

\textbf{Transparency} means that users must be aware of how, when and by whom data about them is collected. This is the most important characteristic of privacy prevention in mobile applications, as many free apps include third party advertising libraries that can collect users' data silently~\cite{sutton2001supporting,vallina2012breaking}.

\section {\textbf{Synopsis of Mobile Privacy Research}}
\label{S:leakage}

Mobile privacy is often investigated at two levels. The Operating System (OS) and the application level are of interest for those looking at issues such as privacy models, data flow, privacy source and sinks in operating systems, effectiveness of current privacy solutions and the analysis of users' attitudes towards privacy. Meanwhile sensors and communication research looks into privacy leaks through dedicated sensors on mobile devices, privacy against sensor data inference, and privacy leaks through mobile communication protocols.

Privacy research in mobile computing is facing many challenges. Data leaks by malicious applications, personal data access by ad libraries, the efficacy of operating systems in protecting data, preventing inference of personal information through mobile sensors, dangers to anonymity of people around the device, awareness and understanding of users for privacy are considered main topics in mobile privacy research. As mobile devices are getting smarter users can now enjoy a diverse range of applications and services on their devices, Mobile social networking, social and location-based recommendations, mobile e-commerce, mobile health, and mobile cloud services are few of major applications, which have become popular among mobile users. These applications access on-device resources to deliver the required services to the users. These applications can work simultaneously in independent or co-operative manner on a mobile device. Diversity in types of services and their mutual co-operation have allowed complex privacy leakages. This in turn has led to poor privacy protection methods and modals. In this section we discuss these privacy leakages and the research carried out in aid of understanding these leakages.

\subsection {Mobile Applications}
All advanced mobile OSs -  \textbf{Android}~\cite{Andrd},  \textbf{Window's phone}~\cite{winphone} and  \textbf{iOS}~\cite{iphoneos}- provide rich software development kits(SDKs), which enables application developers and businesses to implement dynamic applications with ease. These applications can then be executed on these OSs by running the dedicated execution cycle. Moreover, these OSs also provide a permissions model to negotiate the applications' access to personal data. Every request for permission is usually linked to some unique private data. If any application requires the accessing of data, these permission models usually ask permission from the users. However, once permission is granted, the application can then access data associated to that permission for ever. Besides obvious similarities, the implementation of permission models and executions cycles in Android and iOS are different~\cite{book2013privacy}.  

\subsubsection {Android}
Android is a Linux based OS, its applications are written in Java and compile to a custom byte code known as Dalvik byte-code. Each application is executed within the Dalvik virtual machines (DVM) interpreter instance. Each instance is executed as a unique user identity to isolate applications within the Linux platform subsystem. Applications on the same device can communicate by sending parcels via the Binder Inter Process Communication(IPC) mechanism~\cite{Andrdos}. 
 
In Android, all the requests to access sensitive data need to be explicitly included in the application configuration file, at the time of development. During the installation of the application, users have to evaluate any requests and grant the corresponding permissions to continue using the application~\cite{holavanalli2013flow}.

\subsubsection {iOS}
iOS is built on the open source XNU kernel. Applications are written in objective C, and apps are loaded directly by the kernel level loader. The loader interprets the binary and  loads its text and data
segments, and jumps to the app entry point~\cite{iphoneossdk}. 
 
Additionally unlike Android, in iOS there is no concept of explicit permissions and requests. At the time of installation users are not asked for any permission. However, while using, if an application wants to access any personal data, users are asked permissions. In this case users can grant or reject few particular permissions and continue using the application~\cite{HowToGeek12}. These applications are published on dedicated `application stores', from which general users can download and install them on their devices. Developers can earn healthy revenue from their applications as recent reports have estimated billions of Dollars as revenue from these markets~\cite{apprpt2014}. However, due to limited auditing on these markets, applications from them cannot be fully trusted.

Despite privacy controls, it has been found that third party applications can still access and leak personal data without the consent of the users~\cite{egele2011pios}. In addition, a very large proportion of applications on ``application stores" are free of cost. Usually, developers earn a profit from their free applications by including third party advertisement libraries. These libraries can, in turn, access the personal data of users in a hidden manner~\cite{Agarwal:2013:PDM:2462456.2464460}.Privacy research in mobile OSs therefore focus on new privacy leaks and methods to detect them~\cite{egele2011pios,enck2014taintdroid}.

\subsection {Mobile Advertising Libraries}

A large portion of mobile applications are free. To get incentives from these free applications, developers include ads library in their apps. These ads are incorporated inside the applications.  At the run-time of applications, the ad library communicate with an ad network's server to display ads on the user's devices. In doing so these libraries may be sending additional information to the servers without the consent of the users or sometimes even developers. A very detailed explanation of ad libraries and their communication with ad server is provided in~\cite{Grace:2012:UEA:2185448.2185464}.

It has been revealed that these ad libraries contain API calls that can send personal information(users call logs, device IDs, contacts etc) to the ad servers~\cite{Pearce:2}. Additionally, these libraries also gain information that is not required for their purposes. However, this data is significant only when correlated with other user's information. For example few ad libraries send users call log to ad servers, which are not required for target ad displays. Similarly, it is also found that few libraries send information such as phone numbers, SMS service provider and list of installed applications~\cite{Grace:2012:UEA:2185448.2185464}.

\subsection {Mobile Connectivity}

One of most important feature of mobile computing is connectivity. Almost all the applications and services that run on these devices use connectivity to achieve their functionality. These functions may include accessing websites, making calls, communicating with other devices, accessing online services and so on. Modern mobile phones can connect to the internet or other devices by various means. As shown in Table~\ref{table:connect} these technologies greatly differentiate from each other in terms of range, speed of data transfer and main purpose of usage. 
\begin{table}[h]
\begin{tabular*}{\columnwidth}{|p{1.65cm}|p{1.8cm}|p{2.08cm}|l|}
\hline
\hline
{\textbf{Technology}}   & {\textbf{Speed}} & {\textbf{Range}}                                      
& {\textbf{Usage}}                                      
 \\ \hline 
NFC &        424 kbit/s  & Short(\textless 20cm)& P2P comm.  \\ \hline
Bluetooth &  2.1 Mbit/s  & Short(\textless 20ft)& P2P comm. \\ \hline
WiFi &       600 Mbit/s  & Medium(\textless 46m)& Internet  \\ \hline
Cellular &   129 Mbit/s  & Long (\textgreater 40km)& Voice,Internet\\ \hline
\end{tabular*}\\
\caption{Mobile Connectivity Technologies}
\label{table:connect}
\end{table}
Mobile devices are personal to users; therefore, various studies have been contributed by research community, which highlights that personal information can be extracted by analyzing data feeds from these  connectivity technologies in mobile devices~\cite{mulliner2010privacy,xia2013mosaic,kune2012location}. These studies are based on a range of data either collected directly from the network companies(call data records, location histories, network packets) or sniffed data by eavesdropping on the network. 

\subsection {Mobile Sensing}
Sensors enable mobile devices to become aware of the context by providing new dimensions of data. With the advancement in sensor technologies, newer mobile devices are being equipped with new sensors.

Few privacy sensitive sensors in mobile devices and direct data they extract are:
\begin{itemize}[noitemsep,topsep=0pt,parsep=0pt,partopsep=0pt]
	\item
    Camera: {\em  video and images}
    \item
    Biometric: {\em  finger printing, iris imaging}
    \item
    GPS: {\em  location}
    \item
	Accelerometer and Gyroscope: {\em  motion, activity }
 \item
	WiFi, NFC, Bluetooth: {\em  presence of users}
 \item
	Touch: {\em  touch pattern}
\end{itemize}

A growing number of advanced applications are now available to users that use large amount of sensors data from mobile devices to provide a variety of services to the users. The following are a few recent advanced sensor-based applications currently available in App Stores.
\begin{itemize}[noitemsep,topsep=0pt,parsep=0pt,partopsep=0pt]
	\item
   \textbf{CarSafe}~\cite{you2012carsafe} is an application that learns the driving behaviors of users by using the two cameras in a the smart phone.
\item   
    \textbf{Nike+}~\cite{nikeapp} and \textbf{Adidas miCoach}~\cite{malinowski2010adidas} are fitness applications. They track user's activities (route, pace and time) by using GPS and other sensors on mobile devices. This data app also offers personalized coaching to users to improve their running ability.
\item    
    \textbf{StudentLife}~\cite{wang2014studentlife} uses sensing data from the phone to detect the mental health, performance and behavioral trends of the students. 
\item    
   \textbf{NameTage}~\cite{nametagapp} is the first real time facial recognition app for Google Glass. It allows users to capture images from their live video and scan them against photos from social media and dating sites, including more than 450,000 sex offenders. 
\item   
    \textbf{AutoSense}~\cite{ertin2011autosense} is an experimental sensing application, which uses sensors to record physiological data and uses it to understand the psychological state of the user in real time. 
\item 
    \textbf{GasMobile}~\cite{hasenfratz2012participatory} is a paticipatory mobile application for air quality monitoring. It allows users to monitor, visualize and share the information about the air quality.
 \end{itemize}    
    These applications have gone beyond raw location based services and provide a completely new dynamic meaning to the context of the user.

As suggested in~\cite{ghosh2012privacy,jana2013enabling} current privacy controls in the mobile applications are static and therefore they cannot guarantee satisfactory privacy preservation against dynamic context aware applications. Usually these controls ask users to make decisions regarding sensor data. This approach has limitations in a way that they do not provide the user with any information about how the sensor data is captured and used. In summary, the large availability of sensor data and context-aware applications raise new kinds of privacy concerns that are not obvious to the users. Therefore, it is highly important to enhance OSs' privacy models to able to protect the privacy of users against dynamic context aware applications.

\subsection {User attitudes}
Privacy solutions in mobile OSs request users' permission. Many users claim to understand privacy issues in mobile devices, yet studies reveal large amounts of personal data are being released by them through these devices. It has been found that mobile users cannot fully understand and evaluate these permission requests or the contextual value of their personal information~\cite{lin2012expectation}. Owing to this, they are granting permissions to applications that can later harm their privacy~\cite{braunstein2011indirect}. Besides a lack of understanding, other factors also harm mobile users' decisions about the sharing of data. Pedro et al.~\cite{leon2013matters} have listed factors,  which have significant influences on user's behavior towards privacy

\begin{itemize}[noitemsep,topsep=0pt,parsep=0pt,partopsep=0pt]
	\item
   Importance of the type of information;
    \item
   Retention period of data;
    \item
   Usage of collected information;
    \item
	Access and control over collected data;
 \item
	Familiarity and recommendation by friends.
    \end{itemize}

Designing privacy protection solutions, which take into account these all factors, is a challenging task. For example: solution must take into account privacy preferences of individual users to be effective. Application developers have to make sure to provide users access to collected data. The numerous challenges in designing these solutions have led researchers to evaluate users' behavior and expectations. In Section~\ref{S:apps} we discuss some of the issues involved in privacy of applications.

\section {\textbf{Privacy Detection: Methods and Tools}}
\label{S:apps}

The research community has contributed a lot of work to analyze and track how applications leaks private data. A plethora of tools have also developed, which inspect applications for potential privacy leaks. In this section, we first provide preliminary concepts that are important to understand these privacy-monitoring tools. Secondly, we highlight several methodologies followed by them to detect privacy. Finally, we include case studies of some of these tools and privacy leakage detected by them through mobile applications. The aim of this section is to give users an overview of the current and future research in privacy detection tools. Therefore, this section do not provide details about the implementation of these frameworks. Furthermore,~\cite{suarez2013evolution},~\cite{lokhande2014overview}and~\cite{stirparo2013data} have extensively discussed the technical details of these tools and readers may refer to them for further understanding. 

\subsection{Preliminary Concepts}
\textbf{Data Flow Analysis}: Data Flow Analysis(DFA) technique is very popular to track the flow of sensitive information. This technique is dependent on the source and sink of the data. On a higher level, this technique looks for routes between data sources and sinks in mobile OSs as applications run within them. Data sources are sources of sensitive data such as location, file, database and contacts. While data sinks are points that can leak out or leave the mobile device such as the internet or any other mechanism that transmits data out of the system. Any flow of data from source to sink without the user's consent can be classified as privacy leakage~\cite{suarez2013evolution,lokhande2014overview}. Typical sources and sinks for mobile devices as given in~\cite{mann2012framework} are included in Table~\ref{table:source} and Table~\ref{table:sink}.

\textbf{Inter-Component Communication}: Applications are composed of several components, which may include parts of similar or different applications. For example, some components of Android application are;
\begin{itemize}
\item
\textbf{Activity} - controls UI screens
\item
\textbf{Service} - background processes not tied to UI
\item
\textbf{Content Provider} - provides read, write and update operations on app data
\item
\textbf{Broadcast Receiver} - receives messages from Android app framework
\end{itemize}

These components can communicate through objects (like intents). Inter-Component Communication (ICC) can occur either within a single application or across applications~\cite{octeau2013effective}.  

\textbf{Capability Leak}: A capability leak occurs when a malicious app gains access to data by hijacking permissions granted to trustworthy apps without itself requesting them. As mentioned in~\cite{grace2012systematic}, capability leaks can be explicit or implicit. Explicit capability is leaked if the public interface (entry point) of the application exposes capability, which can be exploited or invoked by other unrelated application. While implicit capability is related to internal variables in the setup file of the application. For example, in Android variable `sharedUserId'€ allows apps developed by one developer to have the same User ID. Permissions are granted based on the User IDs of the applications; therefore all applications that share the same User ID gain permission collectively.

\subsection{Privacy Detection Methods}
In the following we have provided some common methods adopted by privacy detection tools. 
\subsubsection{Dynamic Analysis}
Dynamic analysis monitors the behavior of applications to identify privacy leaks as they are executed. In dynamic analysis, the focus is on how the program or application performs on a sensitive input data. By performing DFA through the system, users can be warned about any potential privacy leak through their devices. However, dynamic analysis tools require the actual device or emulator to perform the analysis. Moreover, these tools also have performance overheads as real time analysis of the application is performed~\cite{sarwar2013effectiveness}. 

\begin{figure}
\centering
\captionsetup{justification=centering}
\includegraphics[width=\linewidth]{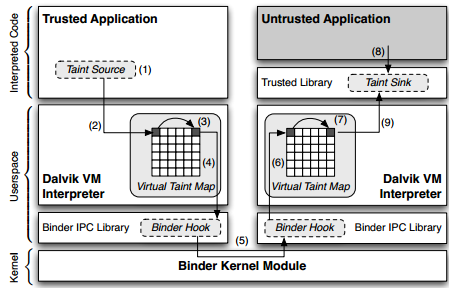}

\caption{TaintDroid Architecture}
\label{figure:Taindroid_Img}

\end{figure}

Figure~\ref{figure:Taindroid_Img} shows the architecture of Taintdroid~\cite{enck2014taintdroid}, a dynamic tool to detect inter application data flow privacy leaks. Information is tainted (1) in the trust application. The taint interface stores specified tag markings in a virtual taint map and also interfaces (2) with Dalvik Vm. The Dalvik VM propagates (3) taint tags as applications use the information included. When a trusted application uses a modified IPC library (4), taint tags are included in the parcel with information and is transferred (5) through kernel transparently to untrustworthy application. At receiving end, the modified IPC library removes taint tags from the parcel and assigns(6) it to all values using the map and Dalvik VM propagates (7) these tags to the application. When an untrustworthy application invokes taint sink library (8), it retrieves tags from the data and reports the event (9).   

\subsubsection{Static analysis}

The static analysis approach tries to cover all possible execution paths of the program. The complete code is statically analyzed without need of its execution, then the control flow graph(CFG) of program is created. The CFG is used to trace the flow of sensitive information from sources to sinks. Modern static analyzers convert programs code into some intermediate representations, which can be effectively processed to generate CFGs. Static analysis takes more time to analyze the program than dynamic analysis as it processes the complete code and all execution paths. However, it has no real time performance overhead, as processing is done statically before the code is actually executed~\cite{lokhande2014overview}. 

\begin{figure}

\includegraphics[width=\linewidth]{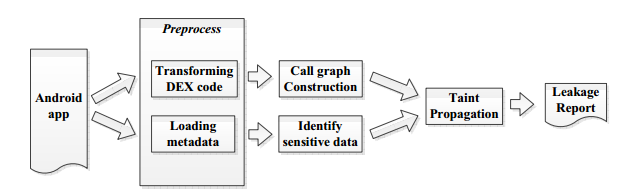}
\centering
\captionsetup{justification=centering}
\caption{LeakMiner Architecture}
\label{figure:LeakMiner_Img}

\end{figure}

In figure~\ref{figure:LeakMiner_Img} the architecture of one static analysis tool LeakMiner~\cite{yang2012leakminer} is given. As mentioned earlier, Android applications are executed in Dalvik byte-code. Therefore this tool first converts byte-code back to Java code and extracts the Meta data of the application, such as permissions, to help identify sensitive data. Using this Meta data, the system then filters relevant API calls. The data flow analysis technique is used to form control flow graph of all instructions and data points dependent on these API calls. If these data points are propagated over the network or logged, a leak path is identified and reported.
\subsubsection{Hybrid}
\begin{figure}

\includegraphics[width=\linewidth]{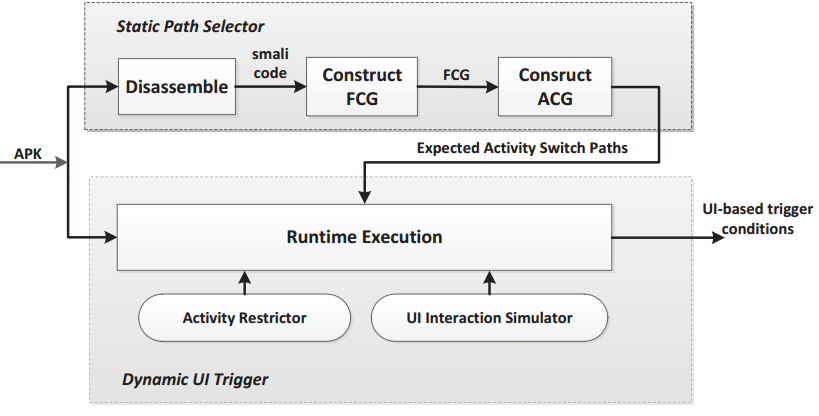}
\centering
\captionsetup{justification=centering}
\caption{SmartDroid Architecture}
\label{figure:SmartDroid_Img}
\end{figure}

A hybrid approach combines static and dynamic analysis to improve the privacy leak detection. In figure~\ref{figure:SmartDroid_Img} an overview of hybrid privacy detection tool, SmartDroid~\cite{zheng2012smartdroid} is shown. At higher level, it implements static path selector, which utilizes static analysis to extract expected activity switching paths by analyzing activity and function CFGs. The dynamic UI trigger then traverses each UI element  to reveal privacy sensitive trigger conditions according to these activity switching paths. 

\subsubsection{Cloud Based Analysis}
Mobile devices are severely restricted in resources, due to which performing privacy detection on them can be problematic. Research community therefore proposes a new cloud based model that devolves privacy detection from mobile devices. 

\begin{figure}

\includegraphics[width=\linewidth]{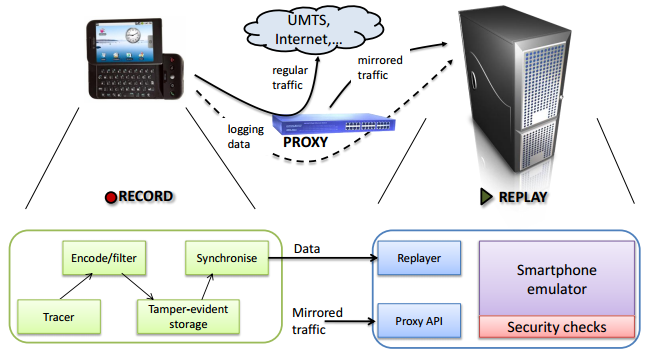}
\centering
\captionsetup{justification=centering}
\caption{Paranoid Android Architecture}
\label{figure:PanDroid_Img}

\end{figure}

An architecture of one such tool named Paranoid Android~\cite{portokalidis2010paranoid} is illustrated in the figure~\ref{figure:PanDroid_Img}. At higher level, it includes running a synchronized replica of the phone on cloud based server. Since server does not have mobile device like constraints, privacy detection analysis that would be too complex to run on mobile devices can be performed. A Tracer, in a mobile device, collects all necessary information required to re-perform mobile application executions. The tracer then transmits this information over encrypted channel to cloud based Replayer that re-executes application in the smart-phone emulator. Afterwards, privacy checks within the emulator can be performed on the server.

\subsubsection{Others}
In the following, we provide a few other methods used by research community to improve the detection of privacy leaks by applications:
\begin{itemize}
\item
\textbf{User's Comments Based} - Rather than analyzing application itself, this method is based on user's comments about the applications. The dataset of users' comments can be collected from official `app stores'. Afterwards, various privacy labels are used to classify privacy related comments~\cite{cenuser}. 
\item
\textbf{Machine Learning Based} - The fundamental principal of dynamic, static and hybrid analysis is to detect the potential flow between sources and sinks. However, most of these methods require fixed list of these sources and sinks as inputs. Recently, research community have proposed an approach that utilize supervised machine learning to automatically generate the list of sources and sinks by analyzing complete application source code~\cite{arzt2013susi}.
\item
\textbf{Mobile Privacy Forensic} - Current privacy detection approaches report leaked private data, but provide limited information about the cause of these leaks. As a result, users are unable to understand the authenticity of the leak. This approach therefore attempts to identify the cause of privacy leak by correlating user actions to leaks~\cite{chan2013case}.  
\item
\textbf{Crowd Sourcing} - In order to determine vulnerabilities in the applications, some of the tools also utilize crowd-sourcing. Crowdroid~\cite{burguera2011crowdroid} uses crowd sourcing to distinguish trustworthy applications from untrustworthy ones having the same names and versions. The use of crowd sourcing provides researchers the behavior traces from different executions of the same application. These traces are then compared to identify a malicious copy of application from the normal one.
\item
\textbf{Privacy Prevention} - Other than monitoring privacy leakages, a few recent tools also provide users the ability to protect against any leakage of private data. This is done by providing fake or bogus data when it is requested by malicious applications. Hornyack et al.~\cite{hornyack2011these} developed the AppFence tool to block sensitive information leakage using the dynamic taint analysis approach. It implements two techniques: data shadowing and exfiltration blocking to restrict applications from leaking sensitive data. Shadowing substitutes shadow data in place of sensitive data to prevent it from exposure and ex-filtration and blocks any network transmission that is carrying sensitive information. Similarly, Mockdroid~\cite{beresford2011mockdroid} and TISSA~\cite{zhou2011taming} also allow users to send fake or mock data to applications. Although providing fake data can affect some part of application functionality, it allows users a trade-off mechanism between privacy and functionality. 

\end{itemize}

\begin{table}[t]

\begin{tabular*}{\columnwidth}{|p{8.42cm}|}
\hline
\hline
{\textbf{Sources}}                                 \\ 
\hline

Location Data: GPS, last base station location, WLAN          \\ \hline
Unique,Identifiers: IMEI, IMSI                                \\ \hline
Authentication,Data: Cashed password data                     \\ \hline
Contact and Calendar,– Contacts, address and schedule         \\ \hline
Call State,– Start and end of incoming call, number of incoming call \\ \hline 
\end{tabular*}
\\
\caption{Sensitive Data Sources}
\label{table:source}

\end{table}

\begin{table}[t]
\begin{tabular*}{\columnwidth}{|p{8.42cm}|}
\hline
\hline
{\textbf{Sinks}}  \\
\hline
 
SMS, Communication: data can be transferred by SMS                            \\ \hline
File Output: Applications can write data to files that are globally readable \\ \hline
Network: Applications can access network by sockets or HTTP                  \\ \hline
Intents, objects: applications can send data objects to other apps           \\ \hline
Content, Resolver Apps can use API to edit shared memory of device          \\  \hline
\end{tabular*}
\\
\caption{Sensitive Data Leaks}
\label{table:sink}
\vspace{-2em}
\end{table}

\subsection{Privacy Leaks Analysis: Case studies}
In this section we have highlighted some of the tools for privacy leak analysis. In addition, we have summarized a list of similar tools in Table~\ref{table:1st_big_table}.
{\bf ScanDAL~\cite{kim2012scandal} :}  This tool performs DFA using static analysis to detect privacy leaks. It converts Dalvik bytecode of Android application packages to a formally defined intermediate language. Dangerous flows are detected using abstract interpretations. ScanDal has analyzed almost 90 free applications out of which 11 were found to leak sensitive data. It has also been found that these applications leaks location data to remote advertisement servers such as AdMob and AdSenseSpec. Moreover, location and phone IMEI is also sent to their application servers by the applications themselves.

{\bf Appintent~\cite{yang2013appintent}:}
Static evaluations were performed on top of 1,000 Android free applications, out of which 248 apps were found to leak some kind of sensitive data. This includes device ID, phone info, location, contacts and SMSs. Researchers also found that many free applications on the Android market still transmit data without the user's awareness; especially mobile social networking applications or applications that integrate ad libraries. On the other hand, it is found that malicious applications also silently leak personal and private data, by combining it with other data that users are aware of. One more interesting finding is about trustworthy applications data logs on devices. Usually these applications log their data onto local logging files in device storage. It is found that sensitive information such as Sim number, device IDs, locations and even contacts are stored temporarily in these files. This logged data can be acquired by malicious applications, hence result in privacy leaks. 

{\bf PCLeaks~\cite{li2014automatically}:}
This tool analyze ICC to detect any potential capability leaks. PCLeaks performed a large scale experiment on 2000 applications. It has been found that large number of applications pose potential capability leaks. Two kinds of component leaks have been found: the potential passive component leak(PPCL), which starts at the entry point of the Android component and ends on the sink. Potential active component leaks(PPAL), starts at the data source and ends at the exiting point of the components. As a result, 43 apps were found to have 143 PACLs, while 147 apps have 843 PPCLs. 

{\bf AndroidLeaks~\cite{grace2012systematic}:}
A very extensive study has been carried out on a large scale on privacy leakage in applications. Almost 7,870 unique leaks have been found from 7,414 Android applications.  The most interesting finding is that 63\% of these leaks were due to the ad codes in the applications. 

{\bf WoodPecker~\cite{grace2012unsafe}:}
As mentioned earlier, Wookpecker tool analyze ICC to detect the capability leaks of the applications. Researchers have used various devices from different manufacturers to detect capability leaks by pre-installed applications. These manufacturers include Google, HTC, Motorola and Samsung. Evaluations have revealed that all pre-loaded apps possess capability leaks. 

{\bf PiOS~\cite{egele2011pios}:}
This tool deals with tracking private data leaks in IOS devices. Analysis has been performed on 1,407 IOS applications. It is found that 657 of these applications include one or more ad or tracking library codes in them. Static analysis looks for all calls to function named `objc msgSend', which is a data transmission function. Through tracking for this function it is found that almost all applications transmit device ID to third party ad libraries and tracking libraries. Additionally, it is also found that applications themselves leak device IDs , location, address books, phone numbers, Safari history and even photos. 

{\bf Kynoid~\cite{schreckling2013kynoid}:} 
This tool enhances Taintdorid and introduces fine grained security permission for individual data items. It is novel in the way that it allows users to specify spatial and temporal constraints on particular data items and restricting the destinations in which they can distribute.

\begin{table*}[htbp!]
\begin{center}

\begin{tabularx}{\textwidth}{|p{2.2cm}|l|p{2cm}|p{1.4cm}|p{0.72cm}|p{8.13cm}|}
\hline
\hline
\textbf{Tools/Frameworks}    & \textbf{Platform} & \textbf{Technique } & \textbf{No. of Tested Apps} & \textbf{Year} & \textbf{Summary } \\ \hline

 Scandal~\cite{kim2012scandal} & Android &	Static Data Flow &	90 & 2011 &	
It is found that 6 applications leak locations to advertisement servers, 5 applications leak locations to analytics server and	1 application leak IMEI to their server.
\\
\hline
PiOS~\cite{egele2011pios} & iOS & Static Data Flow &	1,407 & 2011 &
It is found that 656 applications use ad library code which leak device ID, 195 distinct applications leak Device ID, 36 applications leak GPS location, 5 applications leak address book information and	1 application leak browser history and photo storage
 \\
 \hline
 ProtectMyPrivacy~\cite{agarwal2013protectmyprivacy} &	iOS &	Crowdsourcing &	685 & 2013 &
It is found that 48.43\% applications access Identifier of device, 13.27\% access locations, 6.22\% access contacts and	1.62\% access music library
\\
\hline

AppIntent~\cite{yang2013appintent} &	Android &	Static Data Flow &	1,000 & 2013 &	
It is found that 140 apps have potential data leaks,	26 apps leaks data unintentionally,	24 apps leaks Device ID,	1 app leaks contacts and 1 app leaks SMS
\\
\hline
AndroidLeaks~\cite{gibler2012androidleaks} &	Android & 	Static Data Flow &	25,976 & 2012 &	
Approximately 57,299 leaks are found in applications; 63.51\% leaks are found in ad code. Moreover, 92\% leaks are related to phone data, 5.94\% leaks are of location data and 0.46 and 0.61\% leaks of wiFi and audio
\\
\hline
Woodpecker~\cite{grace2012systematic} & Android &	Capability Leaks & 	953 & 2012 &
Explicit capability(permission) leaks are found in trustworthy applications.

\\
\hline
Mobile Forensics of Privacy Leaks~\cite{stirparo2012mobileak} &	Android & 	Correlate User actions to leaks/ dynamic data flow &	226 & 2012& 	
It is found that 9 different kinds of data is leaked by applications, 34 apps leaks data due to user actions on widgets,	14 leak on start up
and	21 leak data on periodic fraction
\\
\hline
DroidTest~\cite{rumeedroidtest} &	Android &	Dynamic Data flow &	50 & 2013 &	
It is found that most apps leaks model number, subscriber ID, mobile number
\\
\hline
IccTA~\cite{li2014know} &	Android &	Static intra component Analysis & 3,000 & 2014	 &
It is found that 425 applications leaks information directly. Thees leaks are related to device and location data. 

\\
\hline
TISSA~\cite{zhou2011taming} & Android &	Dynamic data Flow &	24 & 2010	&
It is found that 14 apps leak location and	13 leaks device ID.
\\
\hline
PCLeaks~\cite{li2014automatically} &	Android &	Static intra Component analysis &	2,000 & 2010 &	
Nearly 986 component leaks are found. While 534 activity launch leaks are found. Moreover, broadcast injection leaks are 245 and activity hijacking leaks are 110. Additionally, service launch leaks are 64.
\\
\hline
IntentFuzzer~\cite{yang2014intentfuzzer} &	Android & Dynamic Capability leak &	2,183 & 2014 &	
It is found that more than 50\% of applications leak capabilities or permissions related to network state, phone state, location and internet connection.
\\
\hline
Leakminer~\cite{yang2012leakminer} &	Android &	Static Data Flow &	1750 & 2012 &	
It is found that 127 apps leaks device ID, 50 apps leaks phone info, 27 apps leaks Location and 12 apps leaks contacts.
 
\\  \hline
\end{tabularx}
\end{center}
\centering
\caption{Privacy Leak Detection Frameworks}
\label{table:1st_big_table}

\end{table*}

\section {\textbf{Mobile Advertisement}}
\label{S:ads}

In this section we highlight the few privacy concerns unique to mobile ad libraries.
\begin{itemize}

\item
\textbf{Lack of Transparency}
Since mobile ad libraries are incorporated inside host applications, they inherit all the permissions granted to these applications. Therefore, mobile platforms for permission based privacy modals as explained before, are limited in predicting which entity will use these permissions~\cite{Grace:2012:UEA:2185448.2185464}. Moreover, application developers and ad libraries do not promote such practices. Advertisers want information from these permissions to create better user profiles to then better target them with ads~\cite{book2013case,shekhar2012adsplit}.

\item
\textbf{ Undocumented Permissions }
Most ad libraries require the same set of permissions, however few of them also attempt to acquire more privileged undocumented permission. Although none of these permissions are required for the efficient display of ads, many of them can be used to create more complete user profiles. Since these permissions are not documented, developers are not aware of them and hence applications themselves can be considered malicious rather than the library~\cite{stevens2012investigating}.

\item
\textbf{ JavaScript Interface }
Few ad libraries integrate the JavaScript interface, which allows for the dynamic execution of external codes at the run-time. Usually these interfaces expose functionality like making calls, sending SMSs and emailing messages, adding calendar entries, finding locations and making arbitrary network requests. If an adversary is able to inject malicious code into these interfaces then he can perform these operations on any device running particular ad library~\cite{stevens2012investigating}.
\item
\textbf{Tracking}
Multiple applications in the device integrate code from same ad library. Almost all these ad libraries are found to transmit certain kinds of unique device identifiers such as device ID, over the network to their server. These identifiers are particular to the device and can allow malicious ad server to track users across different applications. Moreover, they also provide ground for a network sniffer to track users activities across different ad libraries by mapping the unique identifiers of each device transmitted by different ad libraries~\cite{book2013case}.
\item
\textbf{ Increase in Permissions Usage}
Numbers of studies have revealed that ad libraries use the permissions assigned to applications. However, it has also been found that these ad libraries are increasingly taking advantage of these permissions. In other words research has revealed that there is a steady growth in usage of privacy sensitive permission by ad libraries~\cite{book2013longitudinal}.
\end{itemize}
\section {\textbf{Mobile Connectivity}}
In this section we briefly explain different connectivity technologies available in mobile devices, moreover we also provide case studies of privacy leaks  through these technologies. 

\subsection {Cellular Technology}
\begin{figure}
\centering
\captionsetup{justification=centering}
\includegraphics[width=\linewidth]{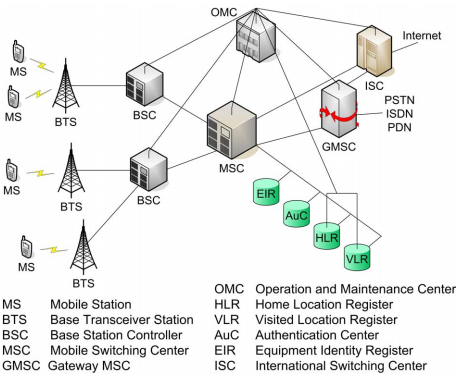}
\caption{Cellular Ecosystem}
\label{figure:Archi_Img}

\end{figure}

Cellular technology allow mobile devices to access the internet plus communicate with other mobile devices through voice communications~\cite{toorani2008solutions}. A typical GSM system and its building blocks are shown in Figure~\ref{figure:Archi_Img}.

\subsubsection {Preliminary Concepts}
Prior to going into the details of privacy leak studies, we provide brief details of core concepts related to cellular mobile networks. 

{\bf Architecture and Components :} As mentioned in~\cite{netarchi} there are 15 main components of this network, however only three of them relate to this survey:
\begin{itemize}[noitemsep,topsep=0pt,parsep=0pt,partopsep=0pt]
\item
	{\bf Mobile Station(MS)} - interacts with the nearest located base stations(BTS) in the cellular network.
\item
	{\bf Base Tower Stations(BTS)} - circulate the data through multiple components of cellular networks to reach their destinations. Interaction between MS and BTS is through wireless protocol that is also called Air Interface.
\item
	{\bf Home location register(HLR) and visitor location register(VLR)} - contain entries for areas of that MS roams in and out and temporary IDs(TMSI) of MS. 
\end{itemize}

{\bf Protocol for Data Flow:} To protect being detected by eavesdropper, mobile phones communicate over cellular networks using temporary identifiers (TMSIs) rather than their long term identifiers (IMSIs). To cater unsuitability, the networks periodically update these identifiers. Other than this, the technical procedure for flow of data on cellular networks is as follows:
\begin{itemize}[noitemsep,topsep=0pt,parsep=0pt,partopsep=0pt]
\item
{\bf Paging Request}- A mobile network attempts to find the MS. The last BTS that has seen the MS, sends a broadcast message with MS's temporary (IMSI) or permanent (TMSI) ID.
\item
{\bf Channeling} - When the MS receives this request, it matches the ID with its own ID. If it matches, MS requests radio resources from its BTS.
\item
{\bf Assignment} - BTS will assign resources to MS, and immediately send an assignment message.
\item
{\bf Paging Response}- MS replies over the resources assigned to it. Later, the protocol allows MS and BTS to set up different parameters and communicate through data. 
\end{itemize}

\begin{figure}
\centering
\captionsetup{justification=centering}
\includegraphics[width=\linewidth]{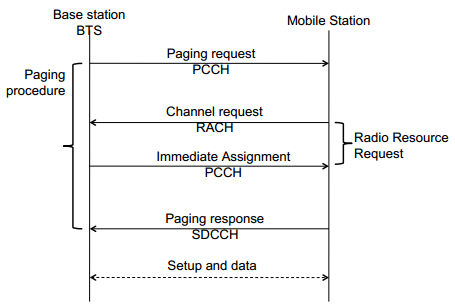}
\centering
\caption{Cellular Protocol}
\label{figure:Protocol_Img}

\end{figure}

In~\cite{kune2012location} researchers have provided a very extensive explanation of cellular protocols and channels associated to them. Moreover, the initial protocols described above are also summarized in Figure~\ref{figure:Protocol_Img}. It should be noted that both paging requests and assignment messages are sent over the broadcast channels with identifiers so that MSs can match their own IDs. Furthermore, BTs send paging requests only for MSs, which are nearby.

\subsubsection{A Surveillance Technology }
As mentioned earlier location data in cellular system is stored in HLR and VLR. HLR is very big database that contains profile information of all the devices on the network. On the other hand VLR is local data repository within BST, it contains profiles of MSs near to BST. This may also includes roaming or users of other network. In addition, HLRs also contains most recent tower ID for each device. This information is kept to efficiently route a data and calls towards particular device. Furthermore, cellular systems also keep track of sectoring (dividing area of each base station into sectors and record most recent sector for each device) and radio signal strength information(RSSI) of each device. One more interesting aspect is that this data is collected and stored for a long period. Moreover , cellular network providers also maintain the mapping between TMSI and IMSI. Clearly, the surveillance ability of the cellular network is evident from these facts. Data from the HLR and VLR alone can be use to track users at the level of BTS. Additionally secotring and  RSSI can make finer tracking possible~\cite{wicker2013cellular}. 

\subsubsection{Privacy Leaks: Case studies}
Like other parts, in this section we again first look at case studies of leaks found by different researchers. Later we summarize these privacy leaks.
\begin{itemize}[noitemsep,topsep=0pt,parsep=0pt,partopsep=0pt]

\item
In~\cite{triukose2012geolocating} Triukose et al. has looked at the feasibility of using IP addresses in cellular data networks to geolocate them. Their data-set includes GPS based data of 29,000 cellular IP addresses in 50 different countries. It is found that mobile networks assign IP address on country level granularity. Furthermore, by experiments, the spatial location of 70\% mobile devices were determined with an error of around 70km. Similarly, in~\cite{eriksson2010learning} researchers showed that by using machines learning clustering techniques geo-location through IP address of mobile devices can be improved. They utilize the naive bayes algorithm that assigns a given IP target to a geographic partition based on a set of measurements associated with that IP target. Through experiments, they were able to determine the location of 96\% mobile devices with an error of 50 km. 
\item
Another privacy leak in cellular networks is investigated by Mulder et al~\cite{de2008identification}. They found that as MSs roam around and register themselves with BTSs, it is possible to identify mobile users from these records and pre- existing location profiles. Moreover, experiments conducted by them identify 80\% of users in the network data-set. 

\item
Similarly, another potential privacy issue has been analyzed by Xia et al.~\cite{xia2013mosaic}. They found that information leaked through networks are fine-grained and also dynamic. It is easier to map users cyber and real world activities by combining data extracted from HTTP headers collected from mobile device traffic and online public profile of users. They were also able to extract shopping behavior and the interest of users by collecting visiting websites names from the cellular network data. 
\item
In addition to tracking users through data recorded by the network providers, the network interface itself can allow silent listening attacks and privacy leaks. A study conducted by Kune et al~\cite{kune2012location} has shown that there is enough information leaking from cellular communication to enable an attacker to perform location tracking on a victim's device.
\end{itemize}
\subsection {WiFi}
WiFi technology has been available for more than a decade. It is a preferred mode to access internet on mobile devices. It is a short range technology and used mainly in public places and houses. Following are the basic elements of WiFi connectivity~\cite{wifiarchi}:
\begin{itemize}
\item
\textbf{Access Point} - an internet connected wireless router, which can connect to mobile devices through WiFi signal.
\item
\textbf{Hot Spot} - an area with accessible WiFi network. This network can be public (allow any mobile device in the vicinity of hot-spot to connect without authentication) or private (requires authentication).
\item
\textbf{Connection Mechanism} - access points broadcast `hello messages' to any device in the vicinity of hot spot. On other side, an individual device can detect these messages and connect to a particular access point. 
\end{itemize}
However, these WiFi networks are not secure. Data is not encrypted, which puts personal data at risk of being sniffed by eavesdropper while using these services. Moreover, a malicious access point can record unencrypted data sent through it~\cite{wifiprivacy}. Security researchers have focused on security threats and solutions in Wi-Fi networks~\cite{najafi2014privacy,cheng2013characterizing}.
\begin{itemize}[noitemsep,topsep=0pt,parsep=0pt,partopsep=0pt]
\item
In~\cite{najafi2014privacy} researchers have revealed the fact that WLAN fingerprints can be used to infer social relation between the users. This can be achieved by measuring similarity between the WiFi fingerprints of the devices. Moreover, since mobile devices broadcast their Wi-Fi information that contain devise ID or MAC address, it is possible for adversaries to actually track locations of devices and users.  
\item
Zero networking is another terminology, which is famous for mobile Wi-Fi networking. The main goal of this networking is to facilitate users to seamlessly connect to devices and services. Devices names are transmitted in this networking protocol to ease the discovery and connection setup of nearby devices. In~\cite{konings2013device} researchers highlighted privacy risks associated with the use of device names in public zero mobile networks, as results revealed that many device names are actually the names of the users. Once users and device have been linked, an individual behavior profile of the user can be created. 
\item
Similar to other research, the researchers in~\cite{cheng2013characterizing} found that user name is the most conspicuous user privacy being leaked. Moreover, it is revealed that users' names can also be detected by analyzing applications, websites and ad content in traffic data through WiFi hot-spots. Since most of the websites and ad libraries store their own resources in files, the websites content can be reached by combining host URL, directory and file name in HTTP protocols.
\item
In another study~\cite{achara2014wifileaks} rather than network data, researchers focus on sensitive information about a device can be accessed by applications through the WiFi interface implemented in these devices. In other words what kind of data applications can be acquired or inferred about devices by calling functions, which are part of a WiFi interface. Their findings are as follows:

\begin{enumerate}
\item
	By having WiFi connection info, applications can get the MAC address of devices, which is a unique ID. Since this ID is permanent, it allows third parties to track users.
\item
	Applications can also learn about the last scanned list of WiFi hot spots around the device. This information includes MAC address, name, signal strength, operating channels and so on. This information can later be geo-locate user positions.
\item
	It is also possible for applications to determine configured network lists on devices. Moreover, by comparing these lists, social relationship between individuals can also be inferred, such as professional, family, interest groups and the like.
\end{enumerate}
\end{itemize}

In summary, researchers have highlighted various privacy concerns related to WiFi connectivity of mobile devices. These concerns include identity exposure and location tracking by using device MAC address and the name of the device. Additionally, social relations among users can also be inferred by comparing their configured networks lists. Moreover, WiFi interfaces are implemented poorly, as is reflected by number of applications that are able to exploit them to access sensitive data.

\section {\textbf{Mobile Sensing}}
\label{S:sensing}

In this section we focus on technologies and applications that utilize mobile sensors. Many of these technologies are mature and are included in our daily life. Moreover, due to advancements in mobile sensors, new technologies have also become part of the mobile computing paradigm. However, studies have also revealed privacy concerns that have been raised by these technologies. In most mobile devices, user-based permissions are associated with each sensor. However, the leakage of sensor data is exacerbated as the public is often unaware of what can be inferred from seemingly harmless data~\cite{mahato2008implicit} and of smart-phone sensing capabilities~\cite{klasnja2009exploring}. Therefore, in this section we first mention users' privacy concerns and limitations in mobile sensor data. Consequently, we explain the privacy risks associated to various technologies based on these sensors. In the section on users behaviors we analyze users' prospects and their understanding of privacy concerns. Here we specifically talk about concerns relating to mobile sensors.

\subsection {Users prospects on sensor data}
Although continuous sensing enables a wide range of applications for users, however privacy concerns of users greatly depend on the type of sensor data collected, for example it has been found that users consider data from GPS sensor to be more sensitive than the data from sensors such as accelerometer and barometer~\cite{raij2011privacy}.

Characteristics of the context in which users are sensed confidentiality requirement of workplace or the perceived vulnerability of the user – also influence users judgment about sensing technology. Specifically reducing or increasing temporal and geographical context effects privacy concerns for a variety of contexts and behaviors~\cite{klasnja2009exploring}.

Another factor is users' perception about the value and importance of functionality that data from particular sensor can enable. If the perceived value does not outweigh the risk of sharing then sensing was rejected. The value of sensor data is perceived by usefulness of functionality to the user and duration data is sensed~\cite{barua2013viewing}.

\subsection{Mobile sensor privacy leaks: Case studies:}

Before going into detail about mobile sensor technologies and privacy leaks associated with them. We provide case studies related to privacy leaks found in raw sensor data.

\begin{itemize}[noitemsep,topsep=0pt,parsep=0pt,partopsep=0pt]
\item A study conducted by Nicholas et al. shows that mobile sensor data exhibits similar sparsity as non-mobile data-sets. Therefore, state-of-the-art de-anonymity techniques can be successfully applied to them. Further, they showed that even with limited background information about a user, an adversary can identify and track the user within an anonymized sensor data set. Moreover, if two sensor data streams are generated by the same individual, then a weekly protected data stream can be used to de anonymize carefully anonymized private data~\cite{lane2012feasibility}.
\item Research by Martina et al. try to find identity pattern of users in their touch sensor usage behavior. Interestingly they were able to identify users with a probability of around 80\%
after just touching ten buttons~\cite{kolly2012personal}.
\item Similarly as mentioned in~\cite{stopczynski2014privacy}, raw sensor data such as GPS and temporal can combined to harm location privacy~\cite{tene2013big}, accelerometers and gyroscopes can be used to track geographical positions or even infer a user's mood~\cite{lane2012feasibility}.
\item Research by Dey et al. shows that due to imperfections in the electro mechanical parts there is diversity in the behavior of the accelerometer. This diversity is not visible from a higher level however if features of these imperfections are extracted, they result in online fingerprints that are enough to identify and track the device~\cite{dey2014accelprint}. In the same way another study also shows that data from an accelerometer can be used to identify different users from the same device~\cite{kwapisz2010cell}.
\end{itemize}

The aforementioned case studies highlight the important fact that data leaks from sensors can be used to infer very personal information about the users. The most important role of these sensors is to provide the context of users to an application or service. We now introduce context awareness. Here we do not survey current research on context-awareness, rather we provide this brief introduction to help readers become more familiar with this area. Interested users may refer to~\cite{musumba2013context} and~\cite{chen2000survey} that provide an in-depth study about context awareness in mobile computing.
 
\subsection{Context awareness:}
Researchers have given various definitions of context. For the purpose of this survey we adopt definition given by Oyomno et al.~\cite{oyomno2009privacy}, which states that €context defines any enriching information about an entity's prevailing situation, including, but not limited to its interactions, attributes and changes to them.€ This means that context aware applications have the ability to use mobile sensors to infer personal information about users such as their environment, activities and even more sensitive attributes like state of mind and adaptability according to user context. As mentioned in \cite{loffler2006quick}, privacy risks in context-aware applications can exist on multiple levels:

{\bf Acquisition:} Acquisition means the collection of context data with the help of a device and its sensors. Each context-aware application uses a few on devices sensors to collect data that in later stages enables them to acquire the context of users. By analyzing the feed of these sensor data, the credentials of the users can be revealed~\cite{gilbert2010toward}.

{\bf Representation:} For the purpose of reusability, data from these sensors is represented in a standard format. This standardization requires data to be clear and easy to understand, so that applications can access this standardized data with ease and prior understanding. However, this reusability sometimes comes at the price of privacy. Due to an easy to understand format it becomes easier for malicious applications to access and alter the data~\cite{chakraborty2011demystifying}.

{\bf Inference:} Inference is the translation of the raw sensor data into information about users' situations, activities, behaviors, and the like. This is the unique feature of context aware applications. Static mobile privacy controls are inadequate in measuring what adversaries can infer from raw sensor data, which gives rise to potential privacy leaks. These inference leaks enable adversaries to gain information about user activities and environments without their consent.

{\bf Transmissions:} Once acquired and represented, the specific contextual data is transmitted to consumers for further processing. It can be a central server or other user devices. Transmission of this contextual information also possesses privacy leaks. An eavesdropper or malicious application can monitor transmission traffic to profile users and their movements. Other details of privacy leaks in transmissions are presented in the network section.

{\bf Utilization:} Context data is then stored by data consumers either for a long time in the repositories or short time inside the devices. Different personally identified information(PII) are removed from this data before releasing it for different usage. However, privacy leaks are inherent in this stored data. Main privacy challenges in this data utilization technique are related to selection of PII, since the availability of a large number of sensor traces can be used to identify users. Additionally, it is assumed that non-PII attributes cannot be linked with an individual's identity. However, the presence of auxiliary information allows adversaries to relate non-PII context data to the identity of individuals~\cite{jagtap2011preserving}.

We have seen that context-aware applications and technologies possess privacy leaks on multiple levels. Mostly privacy leaks specific to context aware applications appear at the acquisition and interpretation of data. Now we introduce technologies based on context awareness and their potential privacy leaks.

\subsection {Location based services}
Presence of GPS and various other common sensors in almost all mobile devices allow applications to adapt themselves to the environment and provide users any useful information that is relevant to the current location of user. This information may include new-targeted advertising, navigation and recommendations. These location aware applications or technology are sub-part of context awareness~\cite{butz2000different}. In addition to direct location data, another concept coupled with location awareness is called proximity awareness, meaning mobile devices are aware of nearby devices. Proximity can be calculated from location data.

There are various methods through which applications can get location information: GPS, mobile phone positioning using network transmission and indoor techniques, which use WiFi and other sources of data~\cite{bergqvist1999location}. A lot of research that deals with privacy leaks in location aware services have been contributed by the research community. Privacy issues proposed in these works are from data acquisition and data interpretation stages of context aware applications. In this section, we include key potential privacy leaks proposed in these research works.

Levijoki et al.~\cite{levijoki2001privacy}, proposed that the most important issue in location aware services is the lack of understanding of whom the apps are providing the information to and for what purposes. Additionally, proximity awareness also raises privacy challenges. Since it allows devices to learn the proximity of each other, this data can also be exploited to track the personal daily routines of users. Minch et al~\cite{minch2004privacy} proposed that location based services do not provide knowledge to users about what data is stored and where. These questions are highly relevant because of the identity issue and the effect of any potential future use of the data. Krumm et al.~\cite{krumm2009survey} shows that other than direct analysis, location data can be used to automatically infer more personal information about the users. These inference leaks may include using data mining techniques to determine the identity of user from data even when it is anonymized~\cite{beresford2003location}. Similarly, the daily routine of users~\cite{gruteser2005anonymity}, clustering location points belonging to the same trajectories~\cite{krumm2006predestination}, predicting the mood of the transportation of data~\cite{krumm2007map}, age, work place and personal habits like smoking and drinking coffee~\cite{kumaraguru2005privacy} can also be inferred from location data.

Freudiger et al. in~\cite{freudiger2012evaluating} showed that Location-Based Services (LBS) providers are able to uniquely identify users and accurately profile them based on some location samples observed from the users. Users with a strong routine face higher privacy risks, especially if their routine does not coincide with that of others. Lee~\cite{leefingerprinting} has shown that the profile of mobile users can be created by analyzing their location tracks. Moreover, these profiles can be used to infer social relationships among the users. Krumm et al. in~\cite{krumm2007inference} was able to identify mobile users on the basis of their location tracks by using a simple algorithm and a free web service. Using GPS mobile data from 172 users, they could find each person's home location with a median error of around 60 meters. Usman et al. in~\cite{iqbal2010privacy} has demonstrated that GPS traces can be used to infer numerous traits about the users by simple algorithms. Jedrzejczyk et al.~\cite{jedrzejczyk2009know} has shown that cross referring location data with publically available information from social network data may lead to full re-identification of users. Moreover, they also demonstrate that by using time stamped mobile location traces the significant locations of a user such as their home, regular patterns in movement, behavior, and location of the place a user works can be identified.

To summarize this section, many privacy leaks have been identified in the literature relating to location aware services. These leaks range from simple acquisition issues of location data like when, who, and where data is collected and stored to inference or interpretation attacks that can result in the extraction of more identifying information from raw location traces.

\subsection {Mobile Augmented Reality}

Augmented reality technology has gained a great deal of acceptance in various applications, for example medical, manufacturing and repair, annotation and visualization, robotics, entertainment and even in the military field. Mobile augmented reality (MAR) has recently become the most discussed and researched field in this study of augmented reality. This is definitely due to the vast availability of mobile and wearable devices. The main theme of this technology is to overlay digital information over the real work that can be viewed from the built-in camera of a device. By doing so MAR has revolutionized the way in which information is presented to the users ~\cite{hollerer2004mobile}.

As suggested in~\cite{azuma1997survey} MAR or AR system usually have the following attributes:

\begin{itemize}[noitemsep,topsep=0pt,parsep=0pt,partopsep=0pt]
\item Combines real and virtual;
\item Is interactive in real time;
\item Is registered in 3-D.
\end{itemize}

Moreover, MAR system depends on following components to perform its functionality:
\begin{itemize}[noitemsep,topsep=0pt,parsep=0pt,partopsep=0pt]
\item A display on which digital content interacting with real world can be shown.
\item Input sensors to collect input information. Camera, microphone, accelerometer sensors are used mostly.
\item Computational and storage power to analyze the input.
\item Network connectivity to keep continuously communicating with application servers.
\end{itemize}

However the field of MAR has also been studied vastly for potential privacy leaks. Here we survey privacy concerns in the field raised by the current literature.

\begin{itemize}
\item \textbf{Surveillance} - The most critical privacy issue for a MAR application is surveillance. MAR applications can record activities of individuals around the device without their consent, due to the `always on' nature of these applications and ability of mobile devices to easily hide. This can raise privacy concerns for those who do not want their normal activities to be recorded~\cite{glassprob}.
\item \textbf{Consent} - Another issues is related to consent. Since MAR applications record data in public places, people around the device have no control over their own data. Current mobile privacy solutions allow mobile users to represent their consent but no solution has been adopted for the consent of the surrounding people~\cite{d2013operating}.
\item  \textbf{Anonymity} - Anonymity is another major privacy issue discussed by the research community. Even though users have public  social network accounts (SNAs) they still want to stay anonymous while in public. However, advancements in face recognition algorithms, may allow these MAR enabled devices to record facial pictures of any person around them and recognize a person's identity by matching his face with his SNA. This issue is recently backed by Acquisiti et al.~\cite{acquisti2011faces} who was able to identify users by taking their photos and matching them with profile data base using face recognition algorithms~\cite{roesner2014security}.
\item \textbf{Inference} - Other issues raised by research community is related to device owners. Since MAR applications can see what user is watching, this data can be misused by malicious services to identify very personal information about users such as daily routine, diet and interests. Moreover, data from MAR applications can also suffer from inference attacks at a larger level than simple location data. For example malicious users can infer relationships, habits, psychological disabilities and so on, about the user of the device~\cite{jana2013enabling}.

\end{itemize}

\subsection {Mobile Health}

Usage of mobile sensors has also been widely accepted in medical health field. Recently mHealth or Mobile Health foundation is also developed by United Nations Organization(UNO). This technology is referred as mHealth. On one hand it enables physicians and doctors to monitor their patients remotely on the other hand it allows patients manage their health in better way with lesser costs. However, privacy issues in mHealth technology are more serious due to very sensitive and personal dealing of health data. Privacy threats in mHealth are discussed in detail by Kotz et al. in~\cite{kotz2011threat}. Major privacy concerns in mHealth technology are as following: 
\begin{itemize}[noitemsep,topsep=0pt,parsep=0pt,partopsep=0pt]
  
  \item \textbf{Identity threats} - There is a risk that patient himself or insider of mHealth system leaks patient's credentials, which allow malicious applications to access personal data related to patient. Moreover, as in the case with location data, even anonymized data can be cross referenced to publicity available data to identify health records of specific patient~\cite{essa2000ubiquitous}.

\item \textbf{Consent} - Another risk that appears in mHealth applications is dependent on patient himself. As patients can control sharing of their data sometimes due to lack of knowledge and worry, they can leak more than required data that can be used by malicious applications to infer personal attributes about the users.

\item \textbf{Disclosure} - Since all mHealth applications deal with very personal and sensitive data so data stored by these applications have risk. Malicious applications can access either through network transmission or direct access to storage, which can results in privacy leaks on very serious level~\cite{huang2010privacy}.

\end{itemize}

\subsection {Mobile Participatory Sensing}
Mobile Participatory Sensing (MPS) leverage the power of millions of personal mobile devices (e.g., smart-phones, wearable devices, sensor-equipped vehicles, etc,), to collect sensing data on large scale without the need to deploy thousands of static sensors. In this paradigm, individuals with sensing and computing devices volunteer to collectively share data and extract information to measure and map phenomena of common interest. Most important feature of MPS is the agreement of nodes to allow their devices to be remotely tasked and routing of the small tasks among the participating mobile nodes to achieve the common goal.

Like other mobile sensing technologies it also suffer from unique privacy issues. A very extensive survey is conducted by Christin et al.~\cite{christin2011survey}, in which they explores the privacy concerns in mobile participatory sensing. Here, we have summarized few of these concerns.

\begin{itemize}
\item 
\textbf{Control of Data} - Although several solutions have been proposed to allow users to control their privacy in particular sensitive data, but due to multiple context requirement of MPS tasks it becomes harder for users to specify policy for each individual data.

\item
\textbf{Tasking} - As mentioned before, one of most interesting feature of MPS is that user's devices can be tasked to sense the data. However, these tasks can have critical threats to users privacy as well. For example, a task in weather sensing MPS application can ask a node to sense weather at particular location. However, it can also leak personal data about the user's mobility and trajectory with respect to time~\cite{kapadia2009opportunistic}. 

Another concern is related to narrow tasking, which means that any malicious user can create the tasks that impose strict limitations on participant attributes or device user is carried. This kind of tasks may reveal private information about the users of node who accept the task. For example, a task may allow the adversary to infer the geographical link between the users~\cite{gilbert2010toward}.

\item
\textbf{Data Delivery} - In addition to tasking, data reporting or delivery can also  pose issues related to user privacy. As users in MPS may volunteer to share their data with central server, however this data can be leaked to malicious users within the network~\cite{kapadia2009opportunistic}.  
 
\item
\textbf{Data Publish} - As mentioned previously, sensor data can be exploited to identify personal information about the users. In MPS, a huge sensor data is collected and stored, if this data is anonymized and published to external entities and organizations, still it can reveal very personal information about the users~\cite{de2013participatory}. For examples researchers have shown that completely anonymous data can be combined with little prior information about the users to reveal complete entity of the user~\cite{narayanan2008robust}.  

\end{itemize}

\section {\textbf{Users Behaviors}}
\label{S:users}

In the ecosystem of mobile computing, users have freedom to install or use certain applications or services. Their decisions to use and share data with applications have high effect on the protection of privacy. Many users claim to understand privacy issues in mobile devices, yet studies reveal large amount of personal data released by them through these devices. This section therefore provides brief overview of user concerns and awareness about privacy. In the Appendix, we have also summarized various users related studies in mobile privacy. Moreover, it describes factors that influence users to make privacy harming decisions.

\subsection {User Concerns and Awareness}
Several studies have been conducted to understand users' point of view about mobile privacy: for instance,~\cite{urban2012mobile} shows that most users are concerned about the protection of personal data in mobile devices. They also oppose practices in which applications collect their personal information like contacts or device ID etc. Moreover they also have concerns about transparency and control of data collection~\cite{minch2004privacy}. Some studies even show that users are more concerned about privacy on their mobile devices than on their laptops. They also prefer to use more critical and personal applications like mobile payments on desktop computers or laptops~\cite{marmasse2000location}. However it is also found that although users are concerned about the privacy leakage in mobile devices, but they have misunderstanding about the sharing of data from their devices. In most of the cases they are not aware of how their data can be used to breach their privacy~\cite{minch2004privacy}.  As a result they may also think privacy as unnecessary abstraction and can make decisions against it. These conclusions are also supported by results in~\cite{felt2012ask},  which shows that majority of users are aware of privacy settings of Facebook mobile application but small proportion of them actually use it.

\subsection {Usability}
It means that the usability and functionality of applications can influence user's understanding about privacy. Studies have shown that users can make compromising decisions against privacy due to their requirement of real time application usage~\cite{muslukhov2012understanding}. They are also lenient about privacy in useful applications that share more data and are strict if applications are useless~\cite{agarwal2013protectmyprivacy}. Which means that they only prefer better privacy if it does not come at the cost of functionality~\cite{good2005stopping}. Moreover, their own expectation about usability and the purpose of why sensitive data is collected also have major impact on their decisions~\cite{lin2012expectation}. Hence it can be concluded that users' decisions are directly affected by their expectations and usability of the applications. If a particular application or service is more usable for them then they are likely to make a tradeoff for privacy.

\subsection {Social Aspects}
Other people also effect users' decisions about privacy. Usually referrals from friends or family, or on-line referrals are the predominant ways by which users discovered new applications for their smartphones. Similarly, popularity, and recommendations from friends also play important role in decision to use the service~\cite{braunstein2011indirect}. It has been found that initially users tend to be more conservative about sharing of personal information with applications. However, as more people around them share data, they become comfortable and relax their privacy policies~\cite{sadeh2009understanding}. In summary, users' decisions, such as ``which application to install?" and ``what data to share?" is highly effected by their social networks.

\subsection {Limitations of privacy solutions}

Poor privacy preserving practices by platforms of mobile devices are also responsible for users' lack of awareness and influence their attitude for making privacy breaching decisions. Current permission models have serious limitations due to which few users read permission requests and even fewer understand them. Human-readable terms displayed before installing an application are vague, confusing, and poorly grouped. This makes it difficult for people, to make informed decisions when installing new software on their mobile devices. Largely, these permissions are ignored and participants instead trust word of mouth, ratings, and Android market reviews~\cite{kelley2012conundrum}.

A study by Kelley et al. also demonstrated that Android users found it difficult to understand the terms and wording of the Android permissions~\cite{kelley2012conundrum}. It is also found that users are not well served by the existing permission architecture~\cite{jung2012short}.  Moreover, they are also not able to make user aware of specific privacy permissions and their importance~\cite{muslukhov2012understanding}~\cite{sadeh2009understanding}. Solutions such as MobiAd~\cite{haddadi:mobiad} prevent direct referral of the user data to advertisers, hence limiting the exposure of the individual to unknown advertisers. However these solutions limit the ability to accurately characterize the success of ad-campaigns and preventing click-fraud. Despite their limitations and challenges, these models can make help users make decisions to protect themselves from obvious data leaks~\cite{haddadi:targeted}.

%% file: conclusion.tex
\section {\textbf{Conclusions and Future Directions}}
\label{S:conc}
\begin{figure*}
\centering
\captionsetup{justification=centering}
\includegraphics[width=\textwidth, height = 18 cm]{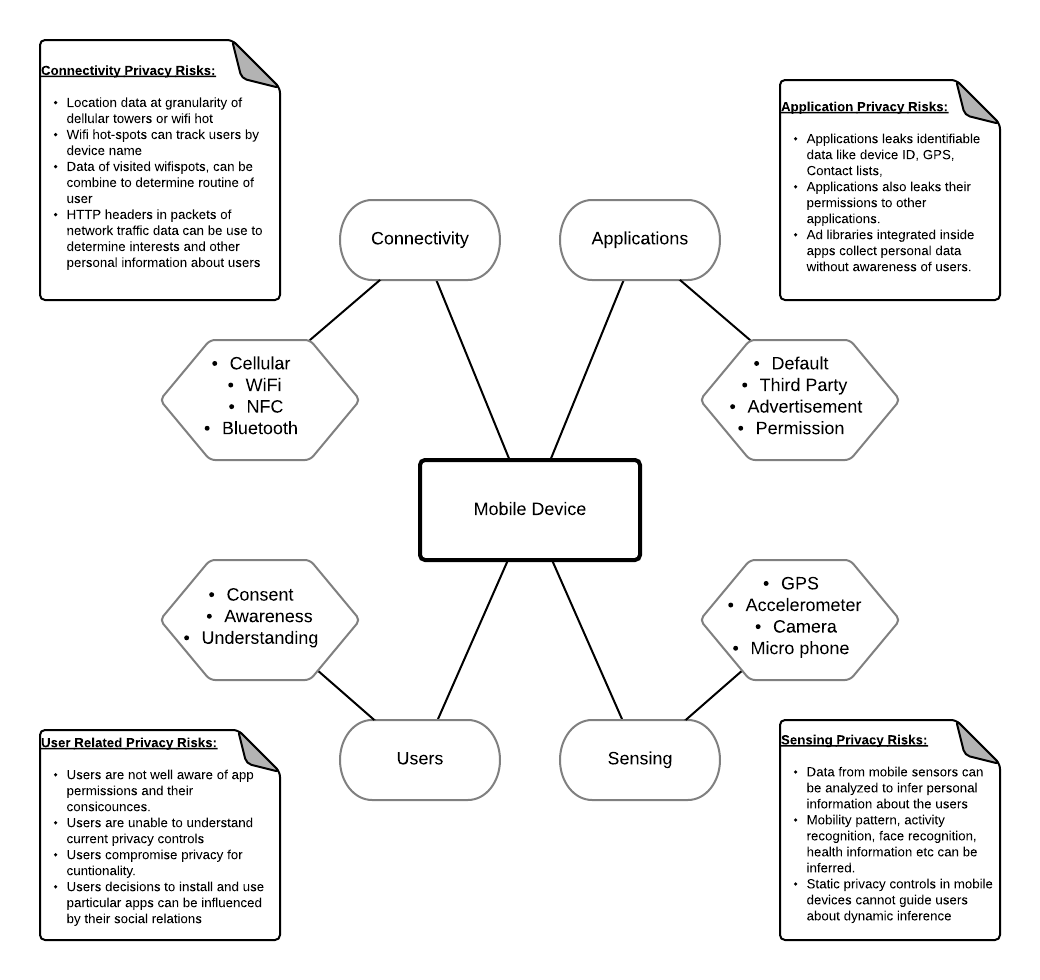}

\caption{Summary of Privacy Research in Mobile Computing}
\label{figure:Overview_Img}
\end{figure*}

In this paper we have provided a comprehensive survey of recent research studies on detection and analysis of privacy leaks in all aspects of mobile computing. We first provided an overview of privacy controls implemented in two main mobile operating systems- Android and iOS. Next we reviewed a few techniques adopted by research communities to detect privacy leaks by mobile applications. We also provided a brief overview of mobile sensing and connectivity. Finally, we demonstrated the user€s' perceptions and views towards keeping their mobile device data private. in order to motivate our readers further about actual privacy leaks, we have also provided case studies of these research studies in each of above-mentioned dimensions of mobile computing. In the mobile applications section, our focus has been on detecting new types of data leaked by applications. We have also presented ongoing research efforts focusing on improving current methods to detect and protect these privacy leaks. 


A large body of research in mobile connectivity has mainly focused on preventing new possible threats which can be performed on current connectivity options such as cellular and WiFi in mobile devices. Additionally researchers are also focusing on making these connectivity protocols more secure. Similarly for mobile sensing, privacy research target potential personal data which can be inferred by mining and combining raw data feeds of various sensors existing on mobile devices. Researchers have also concentrated on privacy preserving techniques for data collection of mobile sensors. Furthermore, research community attempts to find limitations and effectiveness of current privacy controls in mobile computing. Moreover, researcher are also trying to narrow the gap between user preferences, cognitive abilities and privacy controls implementations. In Figure \ref{figure:Overview_Img} we have provided an overview of privacy leaks which have been detected by the research community in the above mentioned research dimensions.

Future research in privacy controls need an extensive comprehensive study to compare different privacy controls systems. The main challenge has been a lack of control modals that takes into account users' cognitive abilities, preferences and limitations to understand the complex privacy options and security flaws. It is clear from the research studies that users are not well served by current privacy controls. Specifically, the human-readable terms displayed before installing an application are at best vague, and at worst confusing, misleading, jargonized, and poorly grouped. This lack of understanding makes it difficult for people, from developers to nontechnical users, to make informed decisions when installing new software on their phones. Largely, the permissions are ignored, with participants instead trusting word of mouth, ratings, and other users' reviews. Hence, there is a clear need for building better mechanisms for preserving privacy in mobile systems~\cite{haddadi2014quantified}. 

Leveraging suggestions of the works surveyed in this paper, we propose a number of recommendations to increase efficiency of privacy controls: (i) Increased transparencies - informing users of the source and destination address when performing sensitive data transfers; (ii) Increased visibility - informing users which applications actually access what data, while differentiating foreground and background applications and exposing hidden features of applications; (iii) Intelligent Suggestions - developing techniques that use machine learning to provide suggestions to users on how to refine their policies based on their own preferences. Research should also be done on recommending users to share or hide data from particular applications because of reputations. Additionally, intelligent privacy leakage control should limit the number to notifications to users while guaranteeing protection. One possible way is to classify the notifications as harmful or harmless, harmless requests should then be granted automatically while others should require user's consent; (iv) Inference Attacks - The control system should inform users about possible inference attacks which can be done with their data. A simple example can illustrate this: if an application asks for multiple sensor feeds at the same time then a user may be notified for potential data inferred from these sensors; (v) Clarity - privacy notifications must be precise and clear. Moreover, visuals warning like privacy widgets can be included in the system that notifies users about any potential sensor data accessed; (vi) Accountability - Once permissions are granted to apps, their behaviors should be analyzed and any anomalous behavior should be reported to the users.

The list presented here is not conclusive. The fast pace in mobile computing, wearables, and IoT will no doubt bring forward a range of new threat models, privacy leakages, and data trade challenges. We hope that this survey will act as a point of reference for future app developers, privacy advocates, and policy makers.

%% file: biographies.tex
\begin{IEEEbiography}
    [{\includegraphics[width=1in,height=1.26in,clip,keepaspectratio]{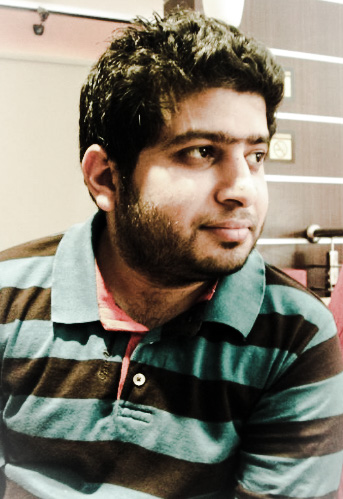}}]{Muhammad Haris} received BS degree in computer system engineering from Ghulam Ishaq Khan Institute of Engineering Sciences and Technology, Pakistan. He is currently working toward a Ph.D. degree at Hong Kong University of Science and Technology, Hong Kong. His research interests include human factors in security system design, mobile security and privacy, human-data interaction and usable privacy. 	
\end{IEEEbiography}
\vspace{-3em}
\begin{IEEEbiography}
    [{\includegraphics[width=1in,height=1.26in,clip,keepaspectratio]{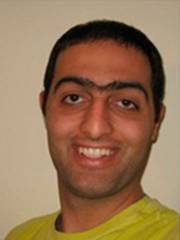}}]{Hamed Haddadi} is the Lecturer in Digital Media at EECS School in Queen Mary University of London and a Research Scientist at Qatar Computing Research Institute. He is interested in Networked Systems \& Social Computing. He enjoys designing and building systems that enable better use of our digital footprint, while respecting users' privacy. He is also broadly interested in sensing applications and Human-Data Interaction. He is currently serving as the Information Services Director for the ACM SIGCOMM Executive Committee.	
\end{IEEEbiography}
\vspace{-3em}
\begin{IEEEbiography}
    [{\includegraphics[width=1in,height=1.26in,clip,keepaspectratio]{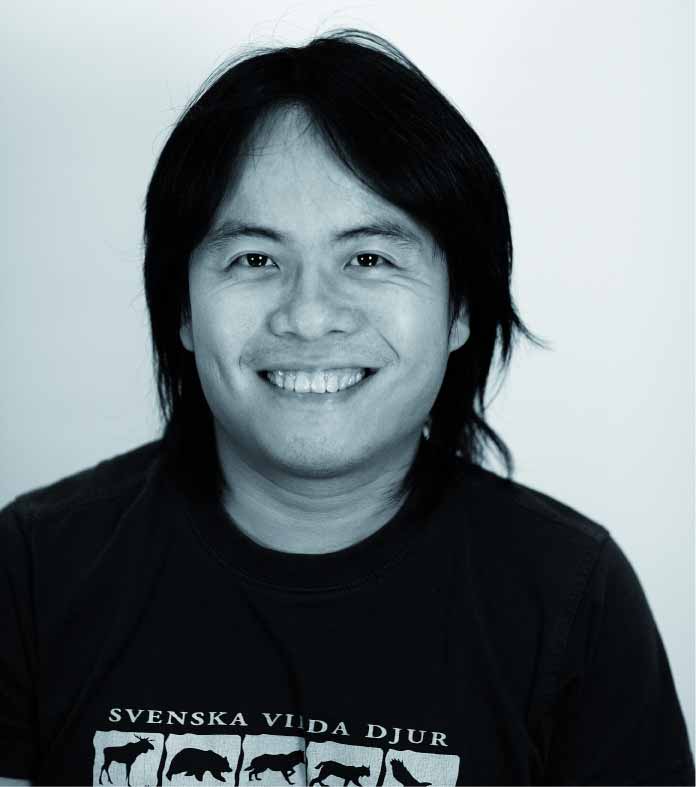}}]{Pan Hui} received his Ph.D degree from Computer Laboratory, University of Cambridge, and earned his MPhil and BEng both from the Department of Electrical and Electronic Engineering, University of Hong Kong. He is currently a faculty member of the Department of Computer Science and Engineering at the Hong Kong University of Science and Technology where he directs the HKUST-DT System and Media Lab. He also serves as a Distinguished Scientist of Telekom Innovation Laboratories (T-labs) Germany and an adjunct Professor of social computing and networking at Aalto University Finland. Before returning to Hong Kong, he has spent several years in T-labs and Intel Research Cambridge. He has published around 150 research papers and has some granted and pending European patents. He has founded and chaired several IEEE/ACM conferences/workshops, and served on the organising and technical program committee of numerous international conferences and workshops including IEEE Infocom, ICNP, SECON, MASS, Globecom, WCNC, ITC, ICWSM and WWW. He is an associate editor for IEEE Transactions on Mobile Computing and IEEE Transactions on Cloud Computing.

\end{IEEEbiography}

%% file: survey.bbl
\begin{thebibliography}{100}
\providecommand{\url}[1]{#1}
\csname url@samestyle\endcsname
\providecommand{\newblock}{\relax}
\providecommand{\bibinfo}[2]{#2}
\providecommand{\BIBentrySTDinterwordspacing}{\spaceskip=0pt\relax}
\providecommand{\BIBentryALTinterwordstretchfactor}{4}
\providecommand{\BIBentryALTinterwordspacing}{\spaceskip=\fontdimen2\font plus
\BIBentryALTinterwordstretchfactor\fontdimen3\font minus
  \fontdimen4\font\relax}
\providecommand{\BIBforeignlanguage}[2]{{%
\expandafter\ifx\csname l@#1\endcsname\relax
\typeout{** WARNING: IEEEtran.bst: No hyphenation pattern has been}%
\typeout{** loaded for the language `#1'. Using the pattern for}%
\typeout{** the default language instead.}%
\else
\language=\csname l@#1\endcsname
\fi
#2}}
\providecommand{\BIBdecl}{\relax}
\BIBdecl

\bibitem{christin2011survey}
D.~Christin, A.~Reinhardt, S.~S. Kanhere, and M.~Hollick, ``A survey on privacy
  in mobile participatory sensing applications,'' \emph{Journal of Systems and
  Software}, vol.~84, no.~11, pp. 1928--1946, 2011.

\bibitem{Clarke:1999:IPC:293411.293475}
\BIBentryALTinterwordspacing
R.~Clarke, ``Internet privacy concerns confirm the case for intervention,''
  \emph{Commun. ACM}, vol.~42, no.~2, pp. 60--67, Feb. 1999. [Online].
  Available: \url{http://doi.acm.org/10.1145/293411.293475}
\BIBentrySTDinterwordspacing

\bibitem{Westin:2003}
A.~F. Westin, ``{S}ocial and {P}olitical {D}imensions of {P}rivacy,''
  \emph{Journal of Social Issues}, vol.~59, no.~2, pp. 431--453, 2003.

\bibitem{Langheinrich02privacyinvasions}
M.~Langheinrich, ``Privacy invasions in ubiquitous computing,'' in
  \emph{WORKSHOP ON SOCIALLY-INFORMED DESIGN OF PRIVACY-ENHANCING SOLUTIONS
  (UBICOMP 2002}.\hskip 1em plus 0.5em minus 0.4em\relax Springer, 2002.

\bibitem{de2013unique}
Y.-A. de~Montjoye, C.~A. Hidalgo, M.~Verleysen, and V.~D. Blondel, ``Unique in
  the crowd: The privacy bounds of human mobility,'' \emph{Scientific reports},
  vol.~3, 2013.

\bibitem{Shmatikov02defininganonymity}
V.~Shmatikov and D.~J.~D. Hughes, ``Defining anonymity and privacy (extended
  abstract),'' 2002.

\bibitem{Introna00privacyand}
L.~D. Introna, ``Privacy and the computer: why we need privacy in the
  information society,'' in \emph{Cyberethics - Social and Moral Issues in the
  Computer Age}.\hskip 1em plus 0.5em minus 0.4em\relax Prometheus Books, 2000,
  pp. 188--199.

\bibitem{sadeh2009understanding}
N.~Sadeh, J.~Hong, L.~Cranor, I.~Fette, P.~Kelley, M.~Prabaker, and J.~Rao,
  ``Understanding and capturing people's privacy policies in a mobile social
  networking application,'' \emph{Personal and Ubiquitous Computing}, vol.~13,
  no.~6, pp. 401--412, 2009.

\bibitem{braunstein2011indirect}
A.~Braunstein, L.~Granka, and J.~Staddon, ``Indirect content privacy surveys:
  measuring privacy without asking about it,'' in \emph{Proceedings of the
  Seventh Symposium on Usable Privacy and Security}.\hskip 1em plus 0.5em minus
  0.4em\relax ACM, 2011, p.~15.

\bibitem{krishnamurthy2008characterizing}
B.~Krishnamurthy and C.~E. Wills, ``Characterizing privacy in online social
  networks,'' in \emph{Proceedings of the first workshop on Online social
  networks}.\hskip 1em plus 0.5em minus 0.4em\relax ACM, 2008, pp. 37--42.

\bibitem{chittaranjan2013mining}
G.~Chittaranjan, J.~Blom, and D.~Gatica-Perez, ``Mining large-scale smartphone
  data for personality studies,'' \emph{Personal and Ubiquitous Computing},
  vol.~17, no.~3, pp. 433--450, 2013.

\bibitem{mohan2008nericell}
P.~Mohan, V.~N. Padmanabhan, and R.~Ramjee, ``Nericell: rich monitoring of road
  and traffic conditions using mobile smartphones,'' in \emph{Proceedings of
  the 6th ACM conference on Embedded network sensor systems}.\hskip 1em plus
  0.5em minus 0.4em\relax ACM, 2008, pp. 323--336.

\bibitem{lu2010jigsaw}
H.~Lu, J.~Yang, Z.~Liu, N.~D. Lane, T.~Choudhury, and A.~T. Campbell, ``The
  jigsaw continuous sensing engine for mobile phone applications,'' in
  \emph{Proceedings of the 8th ACM Conference on Embedded Networked Sensor
  Systems}.\hskip 1em plus 0.5em minus 0.4em\relax ACM, 2010, pp. 71--84.

\bibitem{choudhury2008mobile}
T.~Choudhury, S.~Consolvo, B.~Harrison, J.~Hightower, A.~LaMarca, L.~LeGrand,
  A.~Rahimi, A.~Rea, G.~Bordello, B.~Hemingway \emph{et~al.}, ``The mobile
  sensing platform: An embedded activity recognition system,'' \emph{Pervasive
  Computing, IEEE}, vol.~7, no.~2, pp. 32--41, 2008.

\bibitem{pbd2014}
M.~Birnhack, E.~Toch, and I.~Hadar, ``Privacy mindset, technological mindset,''
  \emph{Jurimetrics}, vol.~55, 2015.

\bibitem{zhou2008brief}
B.~Zhou, J.~Pei, and W.~Luk, ``A brief survey on anonymization techniques for
  privacy preserving publishing of social network data,'' \emph{ACM SIGKDD
  Explorations Newsletter}, vol.~10, no.~2, pp. 12--22, 2008.

\bibitem{ghinita2007fast}
G.~Ghinita, P.~Karras, P.~Kalnis, and N.~Mamoulis, ``Fast data anonymization
  with low information loss,'' in \emph{Proceedings of the 33rd international
  conference on Very large data bases}.\hskip 1em plus 0.5em minus 0.4em\relax
  VLDB Endowment, 2007, pp. 758--769.

\bibitem{ghinita2008anonymization}
G.~Ghinita, Y.~Tao, and P.~Kalnis, ``On the anonymization of sparse
  high-dimensional data,'' in \emph{Data Engineering, 2008. ICDE 2008. IEEE
  24th International Conference on}.\hskip 1em plus 0.5em minus 0.4em\relax
  IEEE, 2008, pp. 715--724.

\bibitem{laurila2012mobile}
J.~K. Laurila, D.~Gatica-Perez, I.~Aad, O.~Bornet, T.-M.-T. Do, O.~Dousse,
  J.~Eberle, M.~Miettinen \emph{et~al.}, ``The mobile data challenge: Big data
  for mobile computing research,'' in \emph{Pervasive Computing}, no.
  EPFL-CONF-192489, 2012.

\bibitem{bugiel2013flexible}
S.~Bugiel, S.~Heuser, and A.-R. Sadeghi, ``Flexible and fine-grained mandatory
  access control on android for diverse security and privacy policies.'' in
  \emph{Usenix security}, 2013, pp. 131--146.

\bibitem{fenske2012biometrics}
J.~Fenske, ``Biometrics in new era of mobile access control,'' \emph{Biometric
  Technology Today}, vol. 2012, no.~9, pp. 9--11, 2012.

\bibitem{sutton2001supporting}
P.~Sutton, R.~Arkins, and B.~Segall, ``Supporting disconnectedness-transparent
  information delivery for mobile and invisible computing,'' in \emph{Cluster
  Computing and the Grid, 2001. Proceedings. First IEEE/ACM International
  Symposium on}.\hskip 1em plus 0.5em minus 0.4em\relax IEEE, 2001, pp.
  277--285.

\bibitem{vallina2012breaking}
N.~Vallina-Rodriguez, J.~Shah, A.~Finamore, Y.~Grunenberger, K.~Papagiannaki,
  H.~Haddadi, and J.~Crowcroft, ``Breaking for commercials: characterizing
  mobile advertising,'' in \emph{Proceedings of the 2012 ACM conference on
  Internet measurement conference}.\hskip 1em plus 0.5em minus 0.4em\relax ACM,
  2012, pp. 343--356.

\bibitem{Andrd}
\BIBentryALTinterwordspacing
A.~Corp. Android. [Online]. Available:
  \url{http://developer.android.com/sdk/index.html}
\BIBentrySTDinterwordspacing

\bibitem{winphone}
\BIBentryALTinterwordspacing
M.~inc. Win phone sdk. [Online]. Available:
  \url{http://dev.windows.com/en-us/develop/download-phone-sdk}
\BIBentrySTDinterwordspacing

\bibitem{iphoneos}
\BIBentryALTinterwordspacing
A.~inc. ios sdk. [Online]. Available:
  \url{https://developer.apple.com/technologies/ios/}
\BIBentrySTDinterwordspacing

\bibitem{book2013privacy}
T.~Book, ``Privacy concerns in android advertising libraries,'' Ph.D.
  dissertation, RICE UNIVERSITY, 2013.

\bibitem{Andrdos}
\BIBentryALTinterwordspacing
A.~Corp. Android sdk. [Online]. Available: \url{http://www.android.com/}
\BIBentrySTDinterwordspacing

\bibitem{holavanalli2013flow}
S.~Holavanalli, D.~Manuel, V.~Nanjundaswamy, B.~Rosenberg, F.~Shen, S.~Y. Ko,
  and L.~Ziarek, ``Flow permissions for android,'' in \emph{Automated Software
  Engineering (ASE), 2013 IEEE/ACM 28th International Conference on}.\hskip 1em
  plus 0.5em minus 0.4em\relax IEEE, 2013, pp. 652--657.

\bibitem{iphoneossdk}
\BIBentryALTinterwordspacing
A.~Inc. ios. [Online]. Available:
  \url{https://developer.apple.com/library/ios/navigation/}
\BIBentrySTDinterwordspacing

\bibitem{HowToGeek12}
HowToGeek, ``ios has app permissions, too: And they’re arguably better than
  android’s,'' 2012.

\bibitem{apprpt2014}
\BIBentryALTinterwordspacing
B.~EVANS. App store revenue. [Online]. Available:
  \url{http://ben-evans.com/benedictevans/2014/7/22/app-store-revenue}
\BIBentrySTDinterwordspacing

\bibitem{egele2011pios}
M.~Egele, C.~Kruegel, E.~Kirda, and G.~Vigna, ``Pios: Detecting privacy leaks
  in ios applications.'' in \emph{NDSS}, 2011.

\bibitem{Agarwal:2013:PDM:2462456.2464460}
\BIBentryALTinterwordspacing
Y.~Agarwal and M.~Hall, ``Protectmyprivacy: Detecting and mitigating privacy
  leaks on ios devices using crowdsourcing,'' in \emph{Proceeding of the 11th
  Annual International Conference on Mobile Systems, Applications, and
  Services}, ser. MobiSys '13.\hskip 1em plus 0.5em minus 0.4em\relax New York,
  NY, USA: ACM, 2013, pp. 97--110. [Online]. Available:
  \url{http://doi.acm.org/10.1145/2462456.2464460}
\BIBentrySTDinterwordspacing

\bibitem{enck2014taintdroid}
W.~Enck, P.~Gilbert, B.-G. Chun, L.~P. Cox, J.~Jung, P.~McDaniel, and A.~N.
  Sheth, ``Taintdroid: an information flow tracking system for real-time
  privacy monitoring on smartphones,'' \emph{Communications of the ACM},
  vol.~57, no.~3, pp. 99--106, 2014.

\bibitem{Grace:2012:UEA:2185448.2185464}
\BIBentryALTinterwordspacing
M.~C. Grace, W.~Zhou, X.~Jiang, and A.-R. Sadeghi, ``Unsafe exposure analysis
  of mobile in-app advertisements,'' in \emph{Proceedings of the Fifth ACM
  Conference on Security and Privacy in Wireless and Mobile Networks}, ser.
  WISEC '12.\hskip 1em plus 0.5em minus 0.4em\relax New York, NY, USA: ACM,
  2012, pp. 101--112. [Online]. Available:
  \url{http://doi.acm.org/10.1145/2185448.2185464}
\BIBentrySTDinterwordspacing

\bibitem{Pearce:2}
\BIBentryALTinterwordspacing
P.~Pearce, A.~P. Felt, G.~Nunez, and D.~Wagner, ``Addroid: Privilege separation
  for applications and advertisers in android,'' in \emph{Proceedings of the
  7th ACM Symposium on Information, Computer and Communications Security}, ser.
  ASIACCS '12.\hskip 1em plus 0.5em minus 0.4em\relax New York, NY, USA: ACM,
  2012, pp. 71--72. [Online]. Available:
  \url{http://doi.acm.org/10.1145/2414456.2414498}
\BIBentrySTDinterwordspacing

\bibitem{mulliner2010privacy}
C.~Mulliner, ``Privacy leaks in mobile phone internet access,'' in
  \emph{Intelligence in Next Generation Networks (ICIN), 2010 14th
  International Conference on}.\hskip 1em plus 0.5em minus 0.4em\relax IEEE,
  2010, pp. 1--6.

\bibitem{xia2013mosaic}
N.~Xia, H.~H. Song, Y.~Liao, M.~Iliofotou, A.~Nucci, Z.-L. Zhang, and
  A.~Kuzmanovic, ``Mosaic: Quantifying privacy leakage in mobile networks,'' in
  \emph{Proceedings of the ACM SIGCOMM 2013 conference on SIGCOMM}.\hskip 1em
  plus 0.5em minus 0.4em\relax ACM, 2013, pp. 279--290.

\bibitem{kune2012location}
D.~F. Kune, J.~Koelndorfer, N.~Hopper, and Y.~Kim, ``Location leaks on the gsm
  air interface,'' \emph{ISOC NDSS (Feb 2012)}, 2012.

\bibitem{you2012carsafe}
C.-W. You, M.~Montes-de Oca, T.~J. Bao, N.~D. Lane, H.~Lu, G.~Cardone,
  L.~Torresani, and A.~T. Campbell, ``Carsafe: a driver safety app that detects
  dangerous driving behavior using dual-cameras on smartphones,'' in
  \emph{Proceedings of the 2012 ACM Conference on Ubiquitous Computing}.\hskip
  1em plus 0.5em minus 0.4em\relax ACM, 2012, pp. 671--672.

\bibitem{nikeapp}
\BIBentryALTinterwordspacing
Nike. Nike+ running. [Online]. Available:
  \url{http://www.nike.com/us/en_us/c/running/nikeplus/gps-app}
\BIBentrySTDinterwordspacing

\bibitem{malinowski2010adidas}
E.~Malinowski, ``Adidas micoach app sets sights square on nike+,'' \emph{Wired
  Magazine}, 2010.

\bibitem{wang2014studentlife}
R.~Wang, F.~Chen, Z.~Chen, T.~Li, G.~Harari, S.~Tignor, X.~Zhou, D.~Ben-Zeev,
  and A.~T. Campbell, ``Studentlife: assessing mental health, academic
  performance and behavioral trends of college students using smartphones,'' in
  \emph{Proceedings of the 2014 ACM International Joint Conference on Pervasive
  and Ubiquitous Computing}.\hskip 1em plus 0.5em minus 0.4em\relax ACM, 2014,
  pp. 3--14.

\bibitem{nametagapp}
\BIBentryALTinterwordspacing
FacialNetwork.com. Nametag application. [Online]. Available:
  \url{http://www.nametag.ws/}
\BIBentrySTDinterwordspacing

\bibitem{ertin2011autosense}
E.~Ertin, N.~Stohs, S.~Kumar, A.~Raij, M.~al'Absi, and S.~Shah, ``Autosense:
  unobtrusively wearable sensor suite for inferring the onset, causality, and
  consequences of stress in the field,'' in \emph{Proceedings of the 9th ACM
  Conference on Embedded Networked Sensor Systems}.\hskip 1em plus 0.5em minus
  0.4em\relax ACM, 2011, pp. 274--287.

\bibitem{hasenfratz2012participatory}
D.~Hasenfratz, O.~Saukh, S.~Sturzenegger, and L.~Thiele, ``Participatory air
  pollution monitoring using smartphones,'' \emph{Mobile Sensing}, 2012.

\bibitem{ghosh2012privacy}
D.~Ghosh, A.~Joshi, T.~Finin, and P.~Jagtap, ``Privacy control in smart phones
  using semantically rich reasoning and context modeling,'' in \emph{Security
  and Privacy Workshops (SPW), 2012 IEEE Symposium on}.\hskip 1em plus 0.5em
  minus 0.4em\relax IEEE, 2012, pp. 82--85.

\bibitem{jana2013enabling}
S.~Jana, D.~Molnar, A.~Moshchuk, A.~M. Dunn, B.~Livshits, H.~J. Wang, and
  E.~Ofek, ``Enabling fine-grained permissions for augmented reality
  applications with recognizers.'' in \emph{USENIX Security}.\hskip 1em plus
  0.5em minus 0.4em\relax Citeseer, 2013, pp. 415--430.

\bibitem{lin2012expectation}
J.~Lin, S.~Amini, J.~I. Hong, N.~Sadeh, J.~Lindqvist, and J.~Zhang,
  ``Expectation and purpose: understanding users' mental models of mobile app
  privacy through crowdsourcing,'' in \emph{Proceedings of the 2012 ACM
  Conference on Ubiquitous Computing}.\hskip 1em plus 0.5em minus 0.4em\relax
  ACM, 2012, pp. 501--510.

\bibitem{leon2013matters}
P.~G. Leon, B.~Ur, Y.~Wang, M.~Sleeper, R.~Balebako, R.~Shay, L.~Bauer,
  M.~Christodorescu, and L.~F. Cranor, ``What matters to users?: factors that
  affect users' willingness to share information with online advertisers,'' in
  \emph{Proceedings of the Ninth Symposium on Usable Privacy and
  Security}.\hskip 1em plus 0.5em minus 0.4em\relax ACM, 2013, p.~7.

\bibitem{suarez2013evolution}
G.~Suarez-Tangil, J.~Tapiador, P.~Peris-Lopez, and A.~Ribagorda, ``Evolution,
  detection and analysis of malware for smart devices,'' 2013.

\bibitem{lokhande2014overview}
B.~Lokhande and S.~Dhavale, ``Overview of information flow tracking techniques
  based on taint analysis for android,'' in \emph{Computing for Sustainable
  Global Development (INDIACom), 2014 International Conference on}.\hskip 1em
  plus 0.5em minus 0.4em\relax IEEE, 2014, pp. 749--753.

\bibitem{stirparo2013data}
P.~Stirparo, I.~N. Fovino, and I.~Kounelis, ``Data-in-use leakages from android
  memory—test and analysis,'' in \emph{Wireless and Mobile Computing,
  Networking and Communications (WiMob), 2013 IEEE 9th International Conference
  on}.\hskip 1em plus 0.5em minus 0.4em\relax IEEE, 2013, pp. 701--708.

\bibitem{mann2012framework}
C.~Mann and A.~Starostin, ``A framework for static detection of privacy leaks
  in android applications,'' in \emph{Proceedings of the 27th Annual ACM
  Symposium on Applied Computing}.\hskip 1em plus 0.5em minus 0.4em\relax ACM,
  2012, pp. 1457--1462.

\bibitem{octeau2013effective}
D.~Octeau, P.~McDaniel, S.~Jha, A.~Bartel, E.~Bodden, J.~Klein, and
  Y.~Le~Traon, ``Effective inter-component communication mapping in android
  with epicc: An essential step towards holistic security analysis,''
  \emph{Effective Inter-Component Communication Mapping in Android with Epicc:
  An Essential Step Towards Holistic Security Analysis}, 2013.

\bibitem{grace2012systematic}
M.~C. Grace, Y.~Zhou, Z.~Wang, and X.~Jiang, ``Systematic detection of
  capability leaks in stock android smartphones.'' in \emph{NDSS}, 2012.

\bibitem{sarwar2013effectiveness}
G.~Sarwar, O.~Mehani, R.~Boreli, and M.~A. Kaafar, ``On the effectiveness of
  dynamic taint analysis for protecting against private information leaks on
  android-based devices.'' in \emph{SECRYPT}, 2013, pp. 461--468.

\bibitem{yang2012leakminer}
Z.~Yang and M.~Yang, ``Leakminer: Detect information leakage on android with
  static taint analysis,'' in \emph{Software Engineering (WCSE), 2012 Third
  World Congress on}.\hskip 1em plus 0.5em minus 0.4em\relax IEEE, 2012, pp.
  101--104.

\bibitem{zheng2012smartdroid}
C.~Zheng, S.~Zhu, S.~Dai, G.~Gu, X.~Gong, X.~Han, and W.~Zou, ``Smartdroid: an
  automatic system for revealing ui-based trigger conditions in android
  applications,'' in \emph{Proceedings of the second ACM workshop on Security
  and privacy in smartphones and mobile devices}.\hskip 1em plus 0.5em minus
  0.4em\relax ACM, 2012, pp. 93--104.

\bibitem{portokalidis2010paranoid}
G.~Portokalidis, P.~Homburg, K.~Anagnostakis, and H.~Bos, ``Paranoid android:
  versatile protection for smartphones,'' in \emph{Proceedings of the 26th
  Annual Computer Security Applications Conference}.\hskip 1em plus 0.5em minus
  0.4em\relax ACM, 2010, pp. 347--356.

\bibitem{cenuser}
L.~Cen, L.~Si, N.~Li, and H.~Jin, ``User comment analysis for android apps and
  cspi detection with comment expansion.''

\bibitem{arzt2013susi}
S.~Arzt, S.~Rasthofer, and E.~Bodden, ``Susi: A tool for the fully automated
  classification and categorization of android sources and sinks,'' 2013.

\bibitem{chan2013case}
J.~J.~K. Chan, K.~W. Tan, L.~Jiang, and R.~K. Balan, ``The case for mobile
  forensics of private data leaks: Towards large-scale user-oriented privacy
  protection.''\hskip 1em plus 0.5em minus 0.4em\relax 4th Asia-Pacific
  Workshop on Systems (APSYS), 2013.

\bibitem{burguera2011crowdroid}
I.~Burguera, U.~Zurutuza, and S.~Nadjm-Tehrani, ``Crowdroid: behavior-based
  malware detection system for android,'' in \emph{Proceedings of the 1st ACM
  workshop on Security and privacy in smartphones and mobile devices}.\hskip
  1em plus 0.5em minus 0.4em\relax ACM, 2011, pp. 15--26.

\bibitem{hornyack2011these}
P.~Hornyack, S.~Han, J.~Jung, S.~Schechter, and D.~Wetherall, ``These aren't
  the droids you're looking for: retrofitting android to protect data from
  imperious applications,'' in \emph{Proceedings of the 18th ACM conference on
  Computer and communications security}.\hskip 1em plus 0.5em minus 0.4em\relax
  ACM, 2011, pp. 639--652.

\bibitem{beresford2011mockdroid}
A.~R. Beresford, A.~Rice, N.~Skehin, and R.~Sohan, ``Mockdroid: trading privacy
  for application functionality on smartphones,'' in \emph{Proceedings of the
  12th Workshop on Mobile Computing Systems and Applications}.\hskip 1em plus
  0.5em minus 0.4em\relax ACM, 2011, pp. 49--54.

\bibitem{zhou2011taming}
Y.~Zhou, X.~Zhang, X.~Jiang, and V.~W. Freeh, ``Taming information-stealing
  smartphone applications (on android),'' in \emph{Trust and Trustworthy
  Computing}.\hskip 1em plus 0.5em minus 0.4em\relax Springer, 2011, pp.
  93--107.

\bibitem{kim2012scandal}
J.~Kim, Y.~Yoon, K.~Yi, J.~Shin, and S.~Center, ``Scandal: Static analyzer for
  detecting privacy leaks in android applications,'' \emph{MoST}, 2012.

\bibitem{yang2013appintent}
Z.~Yang, M.~Yang, Y.~Zhang, G.~Gu, P.~Ning, and X.~S. Wang, ``Appintent:
  Analyzing sensitive data transmission in android for privacy leakage
  detection,'' in \emph{Proceedings of the 2013 ACM SIGSAC conference on
  Computer \& communications security}.\hskip 1em plus 0.5em minus 0.4em\relax
  ACM, 2013, pp. 1043--1054.

\bibitem{li2014automatically}
L.~Li, A.~Bartel, J.~Klein, and Y.~Le~Traon, ``Automatically exploiting
  potential component leaks in android applications,'' 2014.

\bibitem{grace2012unsafe}
M.~C. Grace, W.~Zhou, X.~Jiang, and A.-R. Sadeghi, ``Unsafe exposure analysis
  of mobile in-app advertisements,'' in \emph{Proceedings of the fifth ACM
  conference on Security and Privacy in Wireless and Mobile Networks}.\hskip
  1em plus 0.5em minus 0.4em\relax ACM, 2012, pp. 101--112.

\bibitem{schreckling2013kynoid}
D.~Schreckling, J.~K{\"o}stler, and M.~Schaff, ``Kynoid: real-time enforcement
  of fine-grained, user-defined, and data-centric security policies for
  android,'' \emph{Information Security Technical Report}, vol.~17, no.~3, pp.
  71--80, 2013.

\bibitem{agarwal2013protectmyprivacy}
Y.~Agarwal and M.~Hall, ``Protectmyprivacy: detecting and mitigating privacy
  leaks on ios devices using crowdsourcing,'' in \emph{Proceeding of the 11th
  annual international conference on Mobile systems, applications, and
  services}.\hskip 1em plus 0.5em minus 0.4em\relax ACM, 2013, pp. 97--110.

\bibitem{gibler2012androidleaks}
C.~Gibler, J.~Crussell, J.~Erickson, and H.~Chen, \emph{AndroidLeaks:
  automatically detecting potential privacy leaks in android applications on a
  large scale}.\hskip 1em plus 0.5em minus 0.4em\relax Springer, 2012.

\bibitem{stirparo2012mobileak}
P.~Stirparo and I.~Kounelis, ``The mobileak project: Forensics methodology for
  mobile application privacy assessment,'' in \emph{Internet Technology And
  Secured Transactions, 2012 International Conference for}.\hskip 1em plus
  0.5em minus 0.4em\relax IEEE, 2012, pp. 297--303.

\bibitem{rumeedroidtest}
S.~T.~A. Rumee and D.~Liu, ``Droidtest: Testing android applications for
  leakage of private information.''

\bibitem{li2014know}
L.~Li, A.~Bartel, J.~Klein, Y.~L. Traon, S.~Arzt, S.~Rasthofer, E.~Bodden,
  D.~Octeau, and P.~McDaniel, ``I know what leaked in your pocket: uncovering
  privacy leaks on android apps with static taint analysis,'' \emph{arXiv
  preprint arXiv:1404.7431}, 2014.

\bibitem{yang2014intentfuzzer}
K.~Yang, J.~Zhuge, Y.~Wang, L.~Zhou, and H.~Duan, ``Intentfuzzer: detecting
  capability leaks of android applications,'' in \emph{Proceedings of the 9th
  ACM symposium on Information, computer and communications security}.\hskip
  1em plus 0.5em minus 0.4em\relax ACM, 2014, pp. 531--536.

\bibitem{book2013case}
T.~Book and D.~S. Wallach, ``A case of collusion: A study of the interface
  between ad libraries and their apps,'' in \emph{Proceedings of the Third ACM
  workshop on Security and privacy in smartphones \& mobile devices}.\hskip 1em
  plus 0.5em minus 0.4em\relax ACM, 2013, pp. 79--86.

\bibitem{shekhar2012adsplit}
S.~Shekhar, M.~Dietz, and D.~S. Wallach, ``Adsplit: Separating smartphone
  advertising from applications.'' in \emph{USENIX Security Symposium}, 2012,
  pp. 553--567.

\bibitem{stevens2012investigating}
R.~Stevens, C.~Gibler, J.~Crussell, J.~Erickson, and H.~Chen, ``Investigating
  user privacy in android ad libraries,'' in \emph{Workshop on Mobile Security
  Technologies (MoST)}.\hskip 1em plus 0.5em minus 0.4em\relax Citeseer, 2012.

\bibitem{book2013longitudinal}
T.~Book, A.~Pridgen, and D.~S. Wallach, ``Longitudinal analysis of android ad
  library permissions,'' \emph{arXiv preprint arXiv:1303.0857}, 2013.

\bibitem{toorani2008solutions}
M.~Toorani and A.~Beheshti, ``Solutions to the gsm security weaknesses,'' in
  \emph{Next Generation Mobile Applications, Services and Technologies, 2008.
  NGMAST'08. The Second International Conference on}.\hskip 1em plus 0.5em
  minus 0.4em\relax IEEE, 2008, pp. 576--581.

\bibitem{netarchi}
\BIBentryALTinterwordspacing
3GPP. Network architecture. [Online]. Available:
  \url{http://www.3gpp.org/contact}
\BIBentrySTDinterwordspacing

\bibitem{wicker2013cellular}
S.~B. Wicker, \emph{Cellular Convergence and the Death of Privacy}.\hskip 1em
  plus 0.5em minus 0.4em\relax Oxford University Press, 2013.

\bibitem{triukose2012geolocating}
S.~Triukose, S.~Ardon, A.~Mahanti, and A.~Seth, ``Geolocating ip addresses in
  cellular data networks,'' in \emph{Passive and Active Measurement}.\hskip 1em
  plus 0.5em minus 0.4em\relax Springer, 2012, pp. 158--167.

\bibitem{eriksson2010learning}
B.~Eriksson, P.~Barford, J.~Sommers, and R.~Nowak, ``A learning-based approach
  for ip geolocation,'' in \emph{Passive and Active Measurement}.\hskip 1em
  plus 0.5em minus 0.4em\relax Springer, 2010, pp. 171--180.

\bibitem{de2008identification}
Y.~De~Mulder, G.~Danezis, L.~Batina, and B.~Preneel, ``Identification via
  location-profiling in gsm networks,'' in \emph{Proceedings of the 7th ACM
  workshop on Privacy in the electronic society}.\hskip 1em plus 0.5em minus
  0.4em\relax ACM, 2008, pp. 23--32.

\bibitem{wifiarchi}
\BIBentryALTinterwordspacing
T.~V.~W. Marshall~Brain and B.~Johnson. How do wi-fi hotspots work? [Online].
  Available: \url{http://computer.howstuffworks.com/wireless-network2.htm}
\BIBentrySTDinterwordspacing

\bibitem{wifiprivacy}
\BIBentryALTinterwordspacing
D.~Weedmark. How do wi-fi hotspots work? [Online]. Available:
  \url{http://smallbusiness.chron.com/wifi-hotspots-work-61746.html}
\BIBentrySTDinterwordspacing

\bibitem{najafi2014privacy}
P.~Najafi, A.~Georgiou, D.~Shachneva, and I.~Vlavianos, ``Privacy leaks from
  wi-fi probing1,'' 2014.

\bibitem{cheng2013characterizing}
N.~Cheng, X.~Oscar~Wang, W.~Cheng, P.~Mohapatra, and A.~Seneviratne,
  ``Characterizing privacy leakage of public wifi networks for users on
  travel,'' in \emph{INFOCOM, 2013 Proceedings IEEE}.\hskip 1em plus 0.5em
  minus 0.4em\relax IEEE, 2013, pp. 2769--2777.

\bibitem{konings2013device}
B.~Konings, C.~Bachmaier, F.~Schaub, and M.~Weber, ``Device names in the wild:
  Investigating privacy risks of zero configuration networking,'' in
  \emph{Mobile Data Management (MDM), 2013 IEEE 14th International Conference
  on}, vol.~2.\hskip 1em plus 0.5em minus 0.4em\relax IEEE, 2013, pp. 51--56.

\bibitem{achara2014wifileaks}
J.~P. Achara, M.~Cunche, V.~Roca, A.~Francillon \emph{et~al.}, ``Wifileaks:
  Underestimated privacy implications of the access\_wifi\_state android
  permission,'' 2014.

\bibitem{mahato2008implicit}
H.~Mahato, D.~Kern, P.~Holleis, and A.~Schmidt, ``Implicit personalization of
  public environments using bluetooth,'' in \emph{CHI'08 extended abstracts on
  Human factors in computing systems}.\hskip 1em plus 0.5em minus 0.4em\relax
  ACM, 2008, pp. 3093--3098.

\bibitem{klasnja2009exploring}
P.~Klasnja, S.~Consolvo, T.~Choudhury, R.~Beckwith, and J.~Hightower,
  ``Exploring privacy concerns about personal sensing,'' in \emph{Pervasive
  Computing}.\hskip 1em plus 0.5em minus 0.4em\relax Springer, 2009, pp.
  176--183.

\bibitem{raij2011privacy}
A.~Raij, A.~Ghosh, S.~Kumar, and M.~Srivastava, ``Privacy risks emerging from
  the adoption of innocuous wearable sensors in the mobile environment,'' in
  \emph{Proceedings of the SIGCHI Conference on Human Factors in Computing
  Systems}.\hskip 1em plus 0.5em minus 0.4em\relax ACM, 2011, pp. 11--20.

\bibitem{barua2013viewing}
D.~Barua, J.~Kay, and C.~Paris, ``Viewing and controlling personal sensor data:
  what do users want?'' in \emph{Persuasive Technology}.\hskip 1em plus 0.5em
  minus 0.4em\relax Springer, 2013, pp. 15--26.

\bibitem{lane2012feasibility}
N.~D. Lane, J.~Xie, T.~Moscibroda, and F.~Zhao, ``On the feasibility of user
  de-anonymization from shared mobile sensor data,'' in \emph{Proceedings of
  the Third International Workshop on Sensing Applications on Mobile
  Phones}.\hskip 1em plus 0.5em minus 0.4em\relax ACM, 2012, p.~3.

\bibitem{kolly2012personal}
S.~M. Kolly, R.~Wattenhofer, and S.~Welten, ``A personal touch: recognizing
  users based on touch screen behavior,'' in \emph{Proceedings of the Third
  International Workshop on Sensing Applications on Mobile Phones}.\hskip 1em
  plus 0.5em minus 0.4em\relax ACM, 2012, p.~1.

\bibitem{stopczynski2014privacy}
A.~Stopczynski, R.~Pietri, A.~Pentland, D.~Lazer, and S.~Lehmann, ``Privacy in
  sensor-driven human data collection: A guide for practitioners,'' \emph{arXiv
  preprint arXiv:1403.5299}, 2014.

\bibitem{tene2013big}
O.~Tene and J.~Polonetsky, ``Big data for all: Privacy and user control in the
  age of analytics,'' 2013.

\bibitem{dey2014accelprint}
S.~Dey, N.~Roy, W.~Xu, R.~R. Choudhury, and S.~Nelakuditi, ``Accelprint:
  Imperfections of accelerometers make smartphones trackable,'' in
  \emph{Proceedings of the Network and Distributed System Security Symposium
  (NDSS)}, 2014.

\bibitem{kwapisz2010cell}
J.~R. Kwapisz, G.~M. Weiss, and S.~A. Moore, ``Cell phone-based biometric
  identification,'' in \emph{Biometrics: Theory Applications and Systems
  (BTAS), 2010 Fourth IEEE International Conference on}.\hskip 1em plus 0.5em
  minus 0.4em\relax IEEE, 2010, pp. 1--7.

\bibitem{musumba2013context}
G.~W. Musumba and H.~O. Nyongesa, ``Context awareness in mobile computing: A
  review,'' \emph{International Journal of Machine Learning and Applications},
  vol.~2, no.~1, pp. 5--pages, 2013.

\bibitem{chen2000survey}
G.~Chen, D.~Kotz \emph{et~al.}, ``A survey of context-aware mobile computing
  research,'' Technical Report TR2000-381, Dept. of Computer Science, Dartmouth
  College, Tech. Rep., 2000.

\bibitem{oyomno2009privacy}
W.~Oyomno, P.~J{\"a}ppinen, and E.~Kerttula, ``Privacy implications of
  context-aware services,'' in \emph{Proceedings of the fourth international
  ICST conference on communication system software and middleware}.\hskip 1em
  plus 0.5em minus 0.4em\relax ACM, 2009, p.~17.

\bibitem{loffler2006quick}
T.~L{\"o}ffler, S.~Sigg, S.~Haseloff, and K.~David, ``The quick step to
  foxtrot,'' \emph{Context Awareness for Proactive Systems}, vol. 2006, p. 113,
  2006.

\bibitem{gilbert2010toward}
P.~Gilbert, L.~P. Cox, J.~Jung, and D.~Wetherall, ``Toward trustworthy mobile
  sensing,'' in \emph{Proceedings of the Eleventh Workshop on Mobile Computing
  Systems \& Applications}.\hskip 1em plus 0.5em minus 0.4em\relax ACM, 2010,
  pp. 31--36.

\bibitem{chakraborty2011demystifying}
S.~Chakraborty, H.~Choi, and M.~B. Srivastava, ``Demystifying privacy in
  sensory data: A qoi based approach,'' in \emph{Pervasive Computing and
  Communications Workshops (PERCOM Workshops), 2011 IEEE International
  Conference on}.\hskip 1em plus 0.5em minus 0.4em\relax IEEE, 2011, pp.
  38--43.

\bibitem{jagtap2011preserving}
P.~Jagtap, A.~Joshi, T.~Finin, and L.~Zavala, ``Preserving privacy in
  context-aware systems,'' in \emph{Semantic Computing (ICSC), 2011 Fifth IEEE
  International Conference on}.\hskip 1em plus 0.5em minus 0.4em\relax IEEE,
  2011, pp. 149--153.

\bibitem{butz2000different}
A.~Butz, J.~Baus, A.~Kruger \emph{et~al.}, ``Different views on location
  awareness,'' in \emph{Workshop notes of the ECAI 2000 workshop on Artificial
  Intelligence in Mobile Systems, August 22, 2000, Berlin, Germany}.\hskip 1em
  plus 0.5em minus 0.4em\relax Citeseer, 2000.

\bibitem{bergqvist1999location}
J.~Bergqvist, P.~Dahlberg, H.~Fagrell, and J.~Redstr{\"o}m, ``Location
  awareness and local mobility;-exploring proximity awareness,'' in
  \emph{Proceedings of The Twenty Second IRIS Conference (Information Systems
  Research Seminar In Scandinavia)}, 1999.

\bibitem{levijoki2001privacy}
S.~Levijoki, ``Privacy vs location awareness,'' \emph{Unpublished manuscript,
  Helsinki University of Technology}, 2001.

\bibitem{minch2004privacy}
R.~P. Minch, ``Privacy issues in location-aware mobile devices,'' in
  \emph{System Sciences, 2004. Proceedings of the 37th Annual Hawaii
  International Conference on}.\hskip 1em plus 0.5em minus 0.4em\relax IEEE,
  2004, pp. 10--pp.

\bibitem{krumm2009survey}
J.~Krumm, ``A survey of computational location privacy,'' \emph{Personal and
  Ubiquitous Computing}, vol.~13, no.~6, pp. 391--399, 2009.

\bibitem{beresford2003location}
A.~R. Beresford and F.~Stajano, ``Location privacy in pervasive computing,''
  \emph{Pervasive Computing, IEEE}, vol.~2, no.~1, pp. 46--55, 2003.

\bibitem{gruteser2005anonymity}
M.~Gruteser and B.~Hoh, ``On the anonymity of periodic location samples,'' in
  \emph{Security in Pervasive Computing}.\hskip 1em plus 0.5em minus
  0.4em\relax Springer, 2005, pp. 179--192.

\bibitem{krumm2006predestination}
J.~Krumm and E.~Horvitz, ``Predestination: Inferring destinations from partial
  trajectories,'' in \emph{UbiComp 2006: Ubiquitous Computing}.\hskip 1em plus
  0.5em minus 0.4em\relax Springer, 2006, pp. 243--260.

\bibitem{krumm2007map}
J.~Krumm, E.~Horvitz, and J.~Letchner, ``Map matching with travel time
  constraints,'' SAE Technical Paper, Tech. Rep., 2007.

\bibitem{kumaraguru2005privacy}
P.~Kumaraguru and L.~F. Cranor, ``Privacy indexes: a survey of westin's
  studies,'' 2005.

\bibitem{freudiger2012evaluating}
J.~Freudiger, R.~Shokri, and J.-P. Hubaux, ``Evaluating the privacy risk of
  location-based services,'' in \emph{Financial Cryptography and Data
  Security}.\hskip 1em plus 0.5em minus 0.4em\relax Springer, 2012, pp. 31--46.

\bibitem{leefingerprinting}
G.~M. Lee, ``Fingerprinting and de-anonymizing mobile users.''

\bibitem{krumm2007inference}
J.~Krumm, ``Inference attacks on location tracks,'' in \emph{Pervasive
  Computing}.\hskip 1em plus 0.5em minus 0.4em\relax Springer, 2007, pp.
  127--143.

\bibitem{iqbal2010privacy}
M.~U. Iqbal and S.~Lim, ``Privacy implications of automated gps tracking and
  profiling,'' \emph{Technology and Society Magazine, IEEE}, vol.~29, no.~2,
  pp. 39--46, 2010.

\bibitem{jedrzejczyk2009know}
L.~Jedrzejczyk, B.~A. Price, A.~K. Bandara, and B.~Nuseibeh, ``I know what you
  did last summer: risks of location data leakage in mobile and social
  computing,'' \emph{Department of Computing Faculty of Mathematics, Computing
  and Technology The Open University}, pp. 1744--1986, 2009.

\bibitem{hollerer2004mobile}
T.~H{\"o}llerer and S.~Feiner, ``Mobile augmented reality,''
  \emph{Telegeoinformatics: Location-Based Computing and Services. Taylor and
  Francis Books Ltd., London, UK}, vol.~21, 2004.

\bibitem{azuma1997survey}
R.~T. Azuma \emph{et~al.}, ``A survey of augmented reality,'' \emph{Presence},
  vol.~6, no.~4, pp. 355--385, 1997.

\bibitem{glassprob}
\BIBentryALTinterwordspacing
T.~Claburn. Google glass problems. [Online]. Available:
  \url{http://www.informationweek.com/mobile/7-potential-problems-with-googles-glasses/d/d-id/1102969?}
\BIBentrySTDinterwordspacing

\bibitem{d2013operating}
L.~D’Antoni, A.~Dunn, S.~Jana, T.~Kohno, B.~Livshits, D.~Molnar, A.~Moshchuk,
  E.~Ofek, F.~Roesner, S.~Saponas \emph{et~al.}, ``Operating system support for
  augmented reality applications,'' \emph{Hot Topics in Operating Systems
  (HotOS)}, 2013.

\bibitem{acquisti2011faces}
A.~Acquisti, R.~Gross, and F.~Stutzman, ``Faces of facebook: Privacy in the age
  of augmented reality,'' 2011.

\bibitem{roesner2014security}
F.~Roesner, T.~Kohno, and D.~Molnar, ``Security and privacy for augmented
  reality systems,'' \emph{Communications of the ACM}, vol.~57, no.~4, pp.
  88--96, 2014.

\bibitem{kotz2011threat}
D.~Kotz, ``A threat taxonomy for mhealth privacy.'' in \emph{COMSNETS}, 2011,
  pp. 1--6.

\bibitem{essa2000ubiquitous}
I.~A. Essa, ``Ubiquitous sensing for smart and aware environments,''
  \emph{Personal Communications, IEEE}, vol.~7, no.~5, pp. 47--49, 2000.

\bibitem{huang2010privacy}
X.~Huang, Y.~Jiang, Z.~Liu, T.~Kanter, and T.~Zhang, ``Privacy for mhealth
  presence,'' \emph{International Journal of Next-Generation Networks (IJNGN)},
  vol.~2, no.~4, pp. 33--44, 2010.

\bibitem{kapadia2009opportunistic}
A.~Kapadia, D.~Kotz, and N.~Triandopoulos, ``Opportunistic sensing: Security
  challenges for the new paradigm,'' in \emph{Communication Systems and
  Networks and Workshops, 2009. COMSNETS 2009. First International}.\hskip 1em
  plus 0.5em minus 0.4em\relax IEEE, 2009, pp. 1--10.

\bibitem{de2013participatory}
E.~De~Cristofaro and C.~Soriente, ``Participatory privacy: Enabling privacy in
  participatory sensing,'' \emph{Network, IEEE}, vol.~27, no.~1, pp. 32--36,
  2013.

\bibitem{narayanan2008robust}
A.~Narayanan and V.~Shmatikov, ``Robust de-anonymization of large sparse
  datasets,'' in \emph{Security and Privacy, 2008. SP 2008. IEEE Symposium
  on}.\hskip 1em plus 0.5em minus 0.4em\relax IEEE, 2008, pp. 111--125.

\bibitem{urban2012mobile}
J.~Urban, C.~Hoofnagle, and S.~Li, ``Mobile phones and privacy,'' \emph{UC
  Berkeley Public Law Research Paper}, no. 2103405, 2012.

\bibitem{marmasse2000location}
N.~Marmasse and C.~Schmandt, ``Location-aware information delivery
  withcommotion,'' in \emph{Handheld and Ubiquitous Computing}.\hskip 1em plus
  0.5em minus 0.4em\relax Springer, 2000, pp. 157--171.

\bibitem{felt2012ask}
A.~P. Felt, S.~Egelman, M.~Finifter, D.~Akhawe, D.~Wagner \emph{et~al.}, ``How
  to ask for permission.'' in \emph{HotSec}, 2012.

\bibitem{muslukhov2012understanding}
I.~Muslukhov, Y.~Boshmaf, C.~Kuo, J.~Lester, and K.~Beznosov, ``Understanding
  users' requirements for data protection in smartphones,'' in \emph{Data
  Engineering Workshops (ICDEW), 2012 IEEE 28th International Conference
  on}.\hskip 1em plus 0.5em minus 0.4em\relax IEEE, 2012, pp. 228--235.

\bibitem{good2005stopping}
N.~Good, R.~Dhamija, J.~Grossklags, D.~Thaw, S.~Aronowitz, D.~Mulligan, and
  J.~Konstan, ``Stopping spyware at the gate: a user study of privacy, notice
  and spyware,'' in \emph{Proceedings of the 2005 symposium on Usable privacy
  and security}.\hskip 1em plus 0.5em minus 0.4em\relax ACM, 2005, pp. 43--52.

\bibitem{kelley2012conundrum}
P.~G. Kelley, S.~Consolvo, L.~F. Cranor, J.~Jung, N.~Sadeh, and D.~Wetherall,
  ``A conundrum of permissions: installing applications on an android
  smartphone,'' in \emph{Financial Cryptography and Data Security}.\hskip 1em
  plus 0.5em minus 0.4em\relax Springer, 2012, pp. 68--79.

\bibitem{jung2012short}
J.~Jung, S.~Han, and D.~Wetherall, ``Short paper: enhancing mobile application
  permissions with runtime feedback and constraints,'' in \emph{Proceedings of
  the second ACM workshop on Security and privacy in smartphones and mobile
  devices}.\hskip 1em plus 0.5em minus 0.4em\relax ACM, 2012, pp. 45--50.

\bibitem{haddadi:mobiad}
\BIBentryALTinterwordspacing
H.~Haddadi, P.~Hui, and I.~Brown, ``{MobiAd}: private and scalable mobile
  advertising,'' in \emph{Proceedings of ACM MobiArch}.\hskip 1em plus 0.5em
  minus 0.4em\relax New York, NY, USA: ACM, Sep. 2010, pp. 33--38. [Online].
  Available: \url{http://dx.doi.org/10.1145/1859983.1859993}
\BIBentrySTDinterwordspacing

\bibitem{haddadi:targeted}
H.~Haddadi, P.~Hui, T.~Henderson, and I.~Brown, ``Targeted advertising on the
  handset: Privacy and security challenges,'' in \emph{Pervasive Advertising},
  J.~M\"{u}ller, F.~Alt, and D.~Michelis, Eds.\hskip 1em plus 0.5em minus
  0.4em\relax Springer, Sep. 2011, ch.~6, pp. 119--137.

\bibitem{haddadi2014quantified}
H.~Haddadi and I.~Brown, ``Quantified self and the privacy challenge,''
  \emph{Technology Law Futures}, 2014.

\bibitem{thompson2013s}
C.~Thompson, M.~Johnson, S.~Egelman, D.~Wagner, and J.~King, ``When it's better
  to ask forgiveness than get permission: attribution mechanisms for smartphone
  resources,'' in \emph{Proceedings of the Ninth Symposium on Usable Privacy
  and Security}.\hskip 1em plus 0.5em minus 0.4em\relax ACM, 2013, p.~1.

\bibitem{felt2012android}
A.~P. Felt, E.~Ha, S.~Egelman, A.~Haney, E.~Chin, and D.~Wagner, ``Android
  permissions: User attention, comprehension, and behavior,'' in
  \emph{Proceedings of the Eighth Symposium on Usable Privacy and
  Security}.\hskip 1em plus 0.5em minus 0.4em\relax ACM, 2012, p.~3.

\bibitem{balebako2013little}
R.~Balebako, J.~Jung, W.~Lu, L.~F. Cranor, and C.~Nguyen, ``Little brothers
  watching you: Raising awareness of data leaks on smartphones,'' in
  \emph{Proceedings of the Ninth Symposium on Usable Privacy and
  Security}.\hskip 1em plus 0.5em minus 0.4em\relax ACM, 2013, p.~12.

\bibitem{chin2012measuring}
E.~Chin, A.~P. Felt, V.~Sekar, and D.~Wagner, ``Measuring user confidence in
  smartphone security and privacy,'' in \emph{Proceedings of the Eighth
  Symposium on Usable Privacy and Security}.\hskip 1em plus 0.5em minus
  0.4em\relax ACM, 2012, p.~1.

\bibitem{choe2013nudging}
E.~K. Choe, J.~Jung, B.~Lee, and K.~Fisher, ``Nudging people away from
  privacy-invasive mobile apps through visual framing,'' in
  \emph{Human-Computer Interaction--INTERACT 2013}.\hskip 1em plus 0.5em minus
  0.4em\relax Springer, 2013, pp. 74--91.

\bibitem{king2013come}
J.~King, ``“how come i’m allowing strangers to go through my
  phone?”—smartphones and privacy expectations,'' in \emph{Symposium on
  Usable Privacy and Security (SOUPS)}, 2013.

\bibitem{benenson2013should}
Z.~Benenson and L.~Reinfelder, ``Should the users be informed? on differences
  in risk perception between android and iphone users,'' in \emph{Symposium on
  Usable Privacy and Security (SOUPS)}.\hskip 1em plus 0.5em minus 0.4em\relax
  Citeseer, 2013, pp. 1--2.

\bibitem{barkhuus2003location}
L.~Barkhuus and A.~K. Dey, ``Location-based services for mobile telephony: a
  study of users' privacy concerns.'' in \emph{INTERACT}, vol.~3.\hskip 1em
  plus 0.5em minus 0.4em\relax Citeseer, 2003, pp. 702--712.

\bibitem{kelley2013privacy}
P.~G. Kelley, L.~F. Cranor, and N.~Sadeh, ``Privacy as part of the app
  decision-making process,'' in \emph{Proceedings of the SIGCHI Conference on
  Human Factors in Computing Systems}.\hskip 1em plus 0.5em minus 0.4em\relax
  ACM, 2013, pp. 3393--3402.

\bibitem{shklovski2014leakiness}
I.~Shklovski, S.~D. Mainwaring, H.~H. Sk{\'u}lad{\'o}ttir, and H.~Borgthorsson,
  ``Leakiness and creepiness in app space: perceptions of privacy and mobile
  app use,'' in \emph{Proceedings of the 32nd annual ACM conference on Human
  factors in computing systems}.\hskip 1em plus 0.5em minus 0.4em\relax ACM,
  2014, pp. 2347--2356.

\bibitem{felt2012ve}
A.~P. Felt, S.~Egelman, and D.~Wagner, ``I've got 99 problems, but vibration
  ain't one: a survey of smartphone users' concerns,'' in \emph{Proceedings of
  the second ACM workshop on Security and privacy in smartphones and mobile
  devices}.\hskip 1em plus 0.5em minus 0.4em\relax ACM, 2012, pp. 33--44.

\bibitem{hakkila2005s}
J.~H{\"a}kkil{\"a} and C.~Chatfield, ``'it's like if you opened someone else's
  letter': user perceived privacy and social practices with sms
  communication,'' in \emph{Proceedings of the 7th international conference on
  Human computer interaction with mobile devices \& services}.\hskip 1em plus
  0.5em minus 0.4em\relax ACM, 2005, pp. 219--222.

\bibitem{benisch2011capturing}
M.~Benisch, P.~G. Kelley, N.~Sadeh, and L.~F. Cranor, ``Capturing
  location-privacy preferences: quantifying accuracy and user-burden
  tradeoffs,'' \emph{Personal and Ubiquitous Computing}, vol.~15, no.~7, pp.
  679--694, 2011.

\end{thebibliography}
